\documentclass[a4paper,11pt,openright]{elsarticle}      

\journal{Physica D}

\usepackage{amsmath,amssymb,color}
\usepackage{graphicx}
\usepackage{mathptmx}      
\usepackage{bm}

\newcommand\beq[1]{ \begin{equation}\label{#1} }
\newcommand{\eeq}{ \end{equation} }

\newcommand\beqa[1]{ \begin{eqnarray} \label{#1}}
\newcommand{\eeqa}{ \end{eqnarray} }
\newcommand{\beqano}{ \begin{eqnarray*} }
\newcommand{\eeqano}{ \end{eqnarray*} }
\newcommand\equ[1]{{\rm (\ref{#1})}}

\def\P{{\mathcal P}}
\def\E{{\mathcal E}}

\def\P{{\mathcal P}}
\def\R{{\mathcal R}}
\newcommand{\Z}{\mathbb{Z}}

\def\E{{\mathcal E}}

\def\real{{\mathbb R}}
\def\torus{{\mathbb T}}

\begin{document}

\begin{frontmatter}

\title{Halo orbits around the collinear points of the restricted three-body problem}

\author{Marta Ceccaroni\fnref{fn1}}
\fntext[fn1]{ceccaron@mat.uniroma2.it}
\address{Department of Mathematics, University of Roma Tor Vergata, \\
Via della Ricerca Scientifica, 1 - 00133 Roma}

\author{Alessandra Celletti\fnref{fn2}}
\fntext[fn2]{celletti@mat.uniroma2.it}
\address{Department of Mathematics, University of Roma Tor Vergata,\\
Via della Ricerca Scientifica 1,
00133 Roma (Italy)}

\author{Giuseppe Pucacco\fnref{fn3}}
\fntext[fn3]{Corresponding author: pucacco@roma2.infn.it; telephone/fax: +39 06 72594541.}
\address{
Department of Physics, University of Roma Tor Vergata, \\
Via della Ricerca Scientifica 1,
00133 Roma (Italy)}


\end{frontmatter}

HIGHLIGHTS

$\bullet$ We consider halo orbits around the collinear Lagrangian-Eulerian points.

$\bullet$ An analytical estimate of the bifurcation threshold to halo orbits is obtained.

$\bullet$ The method is based on a normal form adapted to the synchronous resonance.

$\bullet$ A reduction to the central manifold is then performed.

$\bullet$ We make a comparison with available numerical data.

\begin{abstract}
We perform an analytical study of the bifurcation of the halo orbits around the collinear points $L_1$, $L_2$, $L_3$
for the circular, spatial, restricted three--body problem.
Following a standard procedure, we reduce to the center manifold constructing a normal form adapted to the synchronous resonance.
Introducing a detuning, which measures the displacement from the resonance and expanding the energy in series of the detuning, we
are able to evaluate the energy level at which the bifurcation takes place for arbitrary values of the mass ratio.
In most cases, the analytical results thus obtained are in very good agreement with the numerical expectations, providing the
bifurcation threshold with good accuracy.
Care must be taken when dealing with $L_3$ for small values of the mass-ratio between the primaries; in that case, the model of the system is a singular perturbation problem and the normal form
method is not particularly suited to evaluate the bifurcation threshold.
\end{abstract}

\maketitle


\section{Introduction}\label{sec:intro}

In the \sl circular, spatial, restricted three-body problem \rm (hereafter CSR3BP), in the synodic frame, the \sl collinear \rm equilibrium points
 discovered by Euler are located on the line joining the two primaries: $L_1$ lies within them, $L_2$, $L_3$ are outside the interval joining the
primaries. For values of the mass ratio $\mu$ of the primaries corresponding to usual applications like the
barycenter-Sun system (namely, the system describing the interaction between the Earth--Moon barycenter and the Sun) or
the Earth-Moon case, the equilibria $L_1$, $L_2$ are close to the smaller primary, while $L_3$ is quite far on the opposite side of the larger primary. Moreover, two cases are of special interest: $\mu=0$ and $\mu=1/2$.
The limit $\mu \to 0$ can be interpreted as a \sl Hill's problem \rm for $L_1$ and $L_2$, where the two equilibria tend to an equal distance from the smaller primary and the problem is equivalent to let one of the primaries go to infinity as in the
classical lunar theory developed by G.W. Hill in \cite{Hill}; in the case of
$L_3$, again in the limit  $\mu \to 0$, we will speak about a \sl quasi--Kepler problem, \rm
since it is equivalent to a nearly two-body problem in the rotating frame.
On the opposite side of the mass parameter range, namely the case of equal masses, i.e. $\mu=1/2$, we find that the equilibrium $L_1$ is midway from the primaries and $L_2$, $L_3$ are at the same distance from the primaries on each side. The case of such large mass ratio is typically applicable to binary stars or some exotic exo-planetary systems with very large planets.

Overall, in the whole range $\mu \in (0, 1/2]$, we get a very rich dynamical setting with several peculiar phenomena (stability-instability transitions, bifurcations, etc.) characterizing the  non-integrable Hamiltonian system associated to the CSR3BP. As it is well known, the
collinear points are linearly unstable. However, since the seminal paper by C. Conley \cite{conley} on the so--called transit-orbits through $L_1$, much attention has been devoted to the use of the collinear points for space
missions \cite{GLMS}, thanks to the fact that the unstable behavior can be easily controlled for a reasonable time-span.
Moreover, the characteristics of the evolving unstable /stable pathways offer a
structure that can be suitably exploited to design transfer orbits between different regions of the phase space.

On the same ground, the unstable dynamics offers a \sl clean \rm environment free of debris and dust. Indeed, a neighborhood of $L_1$ can be considered a
privileged position to observe the Sun, while $L_2$ is very good to observe the
Universe shielding the Sun through the Earth.

To study the dynamics around these points, numerical
methods provide high accuracy and fast algorithms to follow the
evolution from given initial conditions. However, the analytical
theory gives a deeper insight into the nature of the global
behavior in a neighborhood of these solutions, so to get a comprehensive description of the dynamics in the whole mass range (\cite{JM,CPS,BCCP}).
For example, the phenomena connected with low-order resonances around equilibria play a leading role in shaping the phase-space structure and provide a coarse picture of the global dynamics. Finer details like secondary resonances or heteroclinic intersections are often confined to small portions of the phase-space and usually require dedicated numerical experiments.

The aim of this paper is to present an analytical method to predict the bifurcation thresholds of the \sl halo orbits \rm
around the three collinear points for arbitrary values of the mass ratio: in the present setting this is the most prominent
effect of the resonance in the whole interval between the two extreme cases $\mu=0$ and $\mu=1/2$.
According to Lyapunov's centre theorem, each collinear point generates a pair of one-parameter families of periodic orbits (the nonlinear \sl normal modes\rm),
to which we refer as the planar and the vertical Lyapunov families.
By halo orbit we intend the family of periodic orbits, which arise at the first 1:1 bifurcation from the planar Lyapunov family.
A resonant perturbation theory allows us to
investigate the halo family and to determine the value of the
energy at which the bifurcation from the planar Lyapunov family, namely the \sl horizontal \rm normal mode, takes place \cite{PM14}.
We also construct a normal form to perform the center manifold reduction (see \cite{JM}),
which yields an integrable approximation of the dynamics (compare with \cite{MP14}). The unperturbed linear dynamics on the 2-dimensional center manifold is characterized by almost equal values of the frequencies for all mass ratios. Therefore it is natural to introduce a \sl detuning \rm parameter, which describes the departure from
the exact resonance \cite{Henrard,Vf}. By increasing
the energy, the bifurcations of the 1:1 resonant periodic orbits from the normal modes can be expressed as series expansions in the detuning.

Our results show that for $L_1$ and $L_2$ the prediction of the energy threshold of the
bifurcation is very accurate (up to the fourth decimal digit),
when compared with numerical data available in the literature (see
\cite{GJMS,GM}), even limiting just to a second-order computation; however, we make the effort
of computing higher orders to look for the best agreement (if any) with available data.
Moreover, our strategy allows us to improve previous analytical
approaches based on Lindstedt series \cite{richardson} and to determine first order approximations of the initial conditions for the first bifurcating orbits. In the case of $L_3$, the peculiar nature of the dynamics around it \cite{simo3},
especially when the mass ratio is less than the Earth-Moon value, gives much less accurate results, but our approach is still useful for a qualitative understanding.
In this respect, we provide an explanation for the results concerning $L_3$ in terms of the optimal order of the Birkhoff
normalization procedure applied to a singular perturbation problem.

\vskip.1in

This work is organized as follows. In Section~\ref{sec:collinear}
we present the equations of motion and the location of the collinear points of the CSR3BP.
The corresponding Hamiltonian is diagonalized, normalized and reduced to the center manifold  in
Section~\ref{sec:diag}. In Section~\ref{sec:estimates} we provide analytical formulae for the bifurcation thresholds
at different orders of normalization. In Section~\ref{sec:results} we present the results of our analytical
approach and we compare them with the corresponding numerical values. Section~\ref{sec:conclusion} provides some
conclusions on the results of the present work.

\section{Collinear points in the three-body problem}\label{sec:collinear}

We consider a synodic reference frame centered in the barycenter of the primaries, which are denoted as
$\P_1$, $\P_2$, and rotating with the angular velocity of the primaries. The $X$ axis is set along the line
joining $\P_1$ and $\P_2$, the $Z$ axis along the angular momentum and the $Y$ axis in such a way to have a positively oriented frame.
We normalize the units of measure so that the gravitational constant as well as the sum of the masses of the primaries are unity. Let us rename $\mu$ the mass of the smaller primary; then, with the previous normalization it results that
the larger primary is located\footnote{Notice that with the present convention the equilibrium point $L_2$ is
located to the left of the smaller primary, $L_1$ lies between the primaries and $L_3$ stands at
the right of the larger primary.} at $(\mu,0,0)$, while the
smaller one is at $(\mu-1,0,0)$. The equations of motion of a third small body in the synodic reference frame admit
five equilibrium points discovered by L. Euler and J.-L. Lagrange: the triangular and the collinear points (see, e.g., \cite{Alebook}, \cite{MD}).
The triangular points $L_4$ and $L_5$ are linearly stable whenever $\mu$ is smaller than a threshold, called \sl Routh's value. \rm
On the contrary, the collinear points $L_1$, $L_2$, $L_3$ are shown to be always linearly unstable.

Let us define the kinetic moments $P_X$, $P_Y$, $P_Z$ as
$$
P_X=\dot{X}-Y\ , \qquad  P_Y=\dot{Y}+X\ , \qquad    P_Z=\dot{Z}\ ;
$$
the initial Hamiltonian function describing the motion of the third body is given by
\beq{haminiziale}
H^{(IN)}(P_X,P_Y,P_Z,X,Y,Z)=\frac{1}{2}(P_X^2+P_Y^2+P_Z^2)+YP_X-XP_Y-\frac{1-\mu}{r_1}-\frac{\mu}{r_2}\ ,
\eeq
where $r_1$, $r_2$ denote the distances from the primaries:
$$
r_1=\sqrt{(X-\mu)^2+Y^2+Z^2}\ ,\qquad r_2=\sqrt{(X-\mu+1)^2+Y^2+Z^2}\ .
$$
Let us introduce the scalar function, sometimes called \sl
pseudo-potential \rm (compare with \cite{MD}):
$$
\Omega(X,Y,Z)\equiv \frac{1}{2}(X^2+Y^2)+\frac{1-\mu}{r_1}+\frac{\mu}{r_2}\ ;
$$
then, the equations of motion can be written in compact form as
\beqano
\ddot{X}-2\dot{Y}&=&\frac{\partial \Omega}{\partial X}  \ ,\nonumber\\
\ddot{Y}+2\dot{X}&=&\frac{\partial\Omega}{\partial Y}  \ ,\nonumber\\
\ddot{Z}&=&\frac{\partial\Omega}{\partial Z}\ .
\eeqano
Next we translate the origin so that it coincides with a collinear point; to this end, we determine the distance
$\gamma_j$, $j=1,2,3$, of the collinear equilibria from
the closest primary as the solution of the fifth order Euler's equations (see, e.g., \cite{JM}):
\begin{equation}
\begin{split}
\gamma_1^5-(3-\mu)\gamma_1^4+(3-2\mu)\gamma_1^3-\mu\gamma_1^2+2\mu \gamma_1-\mu&=0  \qquad  \mbox{for}\hspace{1mm} L_1   \ ,\nonumber\\
\gamma_2^5+(3-\mu)\gamma_2^4+(3-2\mu)\gamma_2^3-\mu\gamma_2^2-2\mu \gamma_2-\mu&=0   \qquad   \mbox{for} \hspace{1mm} L_2  \ , \nonumber\\
\gamma_3^5+(2+\mu)\gamma_3^4+(1+2\mu)\gamma_3^3-(1-\mu)\gamma_3^2- 2(1-\mu)\gamma_3-(1-\mu)&=0   \qquad \mbox{for} \hspace{1mm} L_3\ .
\end{split}
\end{equation}
Afterwards, we introduce new coordinates $(x,y,z)$ through the following transformation, which also takes into
account a rescaling of the distances:
$$
X=\mp\gamma_j x+\mu+a\ ,\qquad Y=\mp\gamma_j y\ ,\qquad Z=\gamma_j z\ ,
$$
where the upper signs hold for $L_1$, $L_2$, while the lower signs
are referred to $L_3$; moreover, we set $a=-1+\gamma_1$ for $L_1$,
$a=-1-\gamma_2$ for $L_2$, $a=\gamma_3$ for $L_3$. Denoting by
$P_n=P_n(\chi)$ the Legendre polynomial
 of order $n$ and argument $\chi$, the equations of motion in the new variables can be written in the following
form \cite{JM}, where the pseudo-potential $\Omega$ has been expanded in
terms of the Legendre polynomials:
\beqa{eqmoto2}
\ddot{x}-2\dot{y}-(1+2c_2)x&=&\frac{\partial}{\partial x}\sum_{n\ge 3}c_n(\mu)\rho^n P_n\left(\frac{x}{\rho}\right)\nonumber\\
\ddot{y}+2\dot{x}+(c_2-1)y&=&\frac{\partial}{\partial y}\sum_{n\ge 3}c_n(\mu)\rho^n P_n\left(\frac{x}{\rho}\right)\nonumber\\
\ddot{z}+c_2 z&=&\frac{\partial}{\partial z}\sum_{n\ge
3}c_n(\mu)\rho^n P_n\left(\frac{x}{\rho}\right)\ , \eeqa where
$\rho=\sqrt{x^2+y^2+z^2}$ and where the coefficients $c_n$, $n\geq
2$, are given by the following expressions:
\begin{equation}
\begin{split}
c_n(\mu)=&\frac{1}{\gamma_1^3}\left(\mu +(-1)^n \frac{(1-\mu)\gamma_1^{n+1}}{(1- \gamma_1)^{n+1}}\right)  \qquad\hspace{2mm} \mbox{for} \hspace{2mm} L_1  \ ,\nonumber\\
c_n(\mu)=&\frac{(-1)^n}{\gamma_2^3}\left(\mu   +\frac{(1-\mu)\gamma_2^{n+1}}{(1+ \gamma_2)^{n+1}}\right)  \qquad\hspace{6mm} \mbox{for} \hspace{2mm L_2}  \ , \nonumber\\
c_n(\mu)=&\frac{(-1)^n}{\gamma_3^3}\left(1-\mu +\frac{\mu\gamma_3^{n+1}}{(1+ \gamma_3)^{n+1}}\right)  \qquad \hspace{1mm}\mbox{for} \hspace{2mm} L_3\ .
\end{split}
\end{equation}
Introducing the conjugated momenta $p_x=\dot{x}-y$, $p_y=\dot{y}+x$, $p_z=\dot{z}$, we write the Hamiltonian associated to \equ{eqmoto2} as
\beq{ham1}
H^{(in)}(p_x,p_y,p_z,x,y,z)=\frac{1}{2}\left( p_x^2+p_y^2+p_z^2\right)+yp_x-xp_y- \sum_{n\ge2}c_n(\mu)\rho^n P_n
\left(\frac{x}{\rho}\right)\ .
\eeq
We remark that the relation between $H^{(IN)}$ in \equ{haminiziale} and $H^{(in)}$ in \equ{ham1} is given by (see \cite{GJMS})
\beq{Ephys1}
H^{(IN)}=H^{(in)}\gamma_1^2-{1\over 2}(1-\gamma_1-\mu)^2-{\mu\over \gamma_1}-{{1-\mu}\over {1-\gamma_1}}
\eeq
for $L_1$, by
\beq{Ephys2}
H^{(IN)}=H^{(in)}\gamma_2^2-{1\over 2}(1+\gamma_2-\mu)^2-{\mu\over \gamma_2}-{{1-\mu}\over {1+\gamma_2}}
\eeq
for $L_2$, and by
\beq{Ephys3}
H^{(IN)}=H^{(in)}\gamma_3^2-{1\over 2}(\gamma_3+\mu)^2-{{1-\mu}\over \gamma_3}-{{\mu}\over {1+\gamma_3}}
\eeq
for $L_3$. We also remark that the series at the right hand side of \equ{ham1} is a sum of homogeneous polynomials (with coefficients $c_n(\mu)$), say
$T_n(x,y,z)\equiv\rho^nP_n\left(\frac{x}{\rho}\right)$, which can be iteratively computed by means of the following formulae:
$$
T_0=1\ , \qquad T_1=x\ ,\qquad T_n=\frac{2n-1}{n}xT_{n-1}-\frac{n-1}{n}\rho^2 T_{n-2}\ .
$$


\section{Normalization and center manifold reduction}\label{sec:diag}

In the present section we describe the procedures to construct an integrable approximation of the resonant
dynamics around the collinear points. A straightforward way to achieve this goal consists in performing a resonant
normalization of \equ{ham1}: this is discussed in Sections~\ref{sec:diagonalization} and \ref{norm:form}. A slightly different method has
been adopted in \cite{CPS} on the basis of the approach introduced in \cite{simo2,JM}. This alternative strategy will
be briefly discussed in Section~\ref{sec:indirect}.

\subsection{Diagonalization of the Hamiltonian}\label{sec:diagonalization}
Linearizing \equ{ham1} around the given equilibrium point, we
obtain that the quadratic part of the Hamiltonian is of the form:
\beq{H2}
H_2^{(in)}(p_x,p_y,p_z,x,y,z)=\frac{1}{2}\left(p_x^2+p^2_y\right)+yp_x-xp_y-c_2x^2+
\frac{c_2}{2}y^2+\frac{p_z^2}{2}+\frac{c_2}{2}z^2 \ ,
\eeq
where the coefficient $c_2$ provides the frequency $\omega_z$ of the
$z$-direction, being $\omega_z=\sqrt{c_2}$. We now aim at
diagonalizing \equ{H2} through a standard procedure, that we sketch
here for self--consistency (we refer to \cite{JM,CPS} for full
details). Since the $(p_z,z)$ components are already diagonalized,
let us focus on the remaining variables, for which we write the
equations of motion in the form $\dot{\xi}=J\nabla H_2^{(in)}=M\xi$,
where $\xi\equiv(x,y,p_x,p_y)^T$, $J$ is the symplectic matrix and
$$
M=\begin{pmatrix}
0 & 1 & 1 & 0 \\
-1 & 0&0&1 \\
2c_2 & 0&0&1 \\
0 & -c_2 & -1 &0\\
\end{pmatrix}\ .
$$
The characteristic polynomial associated to $M$ is
$$
p(\lambda)=\lambda^4+(2-c_2)\lambda^2+(1+c_2-2c_2^2)\ ;
$$
the equation $p(\lambda) =0 $ admits the solutions given by the square roots of the quantities $\eta_1$, $\eta_2$, defined as
$$
\eta_1=\frac{c_2-2-\sqrt{9c_2^2-8c_2}}{2}\ , \qquad
\eta_2=\frac{c_2-2+\sqrt{9c_2^2-8c_2}}{2}\ .
$$
Since $c_2>1$,
we have $\eta_1<0$ and $\eta_2>0$, which show that the equilibrium point is of the type
saddle $\times$ center $\times$ center.
Let $\omega_y\equiv\sqrt{-\eta_1}$, $\lambda_x=\sqrt{\eta_2}$; according to \cite{CPS,JM}, we proceed to implement a symplectic change
of variables, such that the quadratic part of the Hamiltonian
is finally diagonalized as
\beq{hqd}
H_2^{(d)}(p_1,p_2,p_3,q_1,q_2,q_3)=\lambda_x q_1p_1+i\omega_y q_2p_2+i\omega_z q_3p_3\ ,
\eeq
where we denote by $(p,q)=(p_1,p_2,p_3,q_1,q_2,q_3)$ the new diagonalizing variables.

\subsection{Resonant normalization}\label{norm:form}
Given the saddle $\times$ center $\times$ center character of the equilibria as shown in Section~\ref{sec:diagonalization},
the center manifold reduction consists in focussing the study to the center directions and in eliminating the hyperbolic
component. Actually, we can perform the center manifold reduction by first constructing the normal form through a suitable canonical transformation, which is obtained by means of Lie series, and then by choosing appropriate initial conditions on the invariant center manifold admitted by the normal form dynamics. We refer to this procedure as the \sl direct method. \rm

We start by expressing the Hamiltonian \equ{ham1} in terms of the diagonalizing variables, so that we obtain the Hamiltonian
\beq{hcd}
H^{(d)}(p_1,p_2,p_3,q_1,q_2,q_3)=\sum_{n\ge 2}H_n^{(d)}(p,q)\ ,
\eeq
where $H_2^{(d)}$ is given by \equ{hqd} and $H_n^{(d)}$ are homogeneous polynomials of degree $n$. Next we proceed to perform a resonant
perturbation theory in the neighborhood of the synchronous resonance $\omega_y=\omega_z$ (see \cite{Alebook, ferraz})
by constructing a canonical transformation, $(p,q) \longrightarrow (P,Q)$,
which conjugates \equ{hcd} to the form:
\beqa{zero}
K^{(NF)}(P_1,P_2,P_3,Q_1,Q_2,Q_3)&=&\lambda_x Q_1P_1+i\omega_y Q_2P_2+i\omega_z Q_3P_3\nonumber\\
&+&\sum_{n=3}^N K^{(NF)}_n(Q_1P_1,P_2,P_3,Q_2,Q_3)+R_{N+1}(P,Q)\ ,\nonumber\\
\eeqa
where the homogeneous polynomials $K^{(NF)}_n$, $n=3,...,N$, are in \sl normal form \rm with respect to the (synchronous)
resonant quadratic part $K^{(NF)}_2=H_2^{(r)}$ with $H_2^{(r)}$ given by
\beq{hqr}
H_2^{(r)}(P_1,P_2,P_3,Q_1,Q_2,Q_3) \equiv \lambda_x Q_1P_1+i\omega_z ( Q_2P_2+  Q_3P_3)
\eeq
and $R_{N+1}(P,Q)$ is a remainder function of degree $N+1$. By \sl normal form \rm we mean that each term up to order $N$ in the series  \equ{zero} satisfies the condition
$$
\{H_2^{(r)},K^{(NF)}_n\} = 0 \ ,
$$
where $\{\cdot,\cdot\}$ denotes the Poisson brackets. The resonant quadratic Hamiltonian in \equ{hqr}, $H_2^{(r)}(P_1,P_2,P_3,Q_1,Q_2,Q_3)$, is obtained from the original quadratic part in \equ{hqd}, expressed in the new variables and modified in order to be resonant in the elliptic components.  We can justify this assumption by observing that, for any $\mu \in (0,1/2]$ the two elliptic frequencies are such that the quantity
$$
\delta \equiv \omega_y-\omega_z\ ,
$$
to which we refer as the \sl detuning, \rm is always a small quantity (in our examples it will be of the order of $10^{-2}$).
The detuning provides a measure of the distance in the frequency from the synchronous
resonance. In this way, even if the unperturbed system is strictly not resonant, we are able to describe
the resonant dynamics of the perturbed system determined by the nonlinear coupling. The detuning parameter will be
used to obtain series expansions of indicators, such as the bifurcation thresholds to halo orbits.

Since the normalization involving the hyperbolic components is a standard Birkhoff normalization, the normal form depends on $Q_1$, $P_1$ only through their product, while the remainder $R_{N+1}(P,Q)$ might depend on $Q_1$, $P_1$ separately.


\subsection{Center manifold reduction}\label{sec:center}
For convenience, we implement the change of variables
\begin{equation}
\begin{cases}
Q_1=\sqrt{I_x}e^{\theta_x} \nonumber\\
Q_2=\sqrt{I_y}(\sin \theta_y -i\cos \theta_y)=-i\sqrt{I_y}e^{i\theta_y} \nonumber\\
Q_3=\sqrt{I_z}(\sin \theta_z -i\cos \theta_z)=-i\sqrt{I_z}e^{i\theta_z} \nonumber\\
P_1=\sqrt{I_x}e^{-\theta_x} \nonumber\\
P_2=\sqrt{I_y}(\cos \theta_y -i\sin \theta_y)=\sqrt{I_y}e^{-i\theta_y}  \nonumber\\
P_3=\sqrt{I_z}(\cos \theta_z -i\sin \theta_z)=\sqrt{I_z}e^{-i\theta_z}\ .\nonumber
\end{cases}
\end{equation}
From the structure of the normal form Hamiltonian \equ{zero} we see that the \sl action \rm variable $I_x=Q_1P_1$ is a constant
of motion, whenever the remainder is neglected. Therefore, given an initial condition $I_x(0)=0$ and neglecting $R_{N+1}$, we obtain an integrable Hamiltonian in two degrees of freedom (hereafter, DOF), which provides the dynamics in the center manifold
up to an approximation of order $N$. Within the center manifold, we describe the motion by the following
2-DOF Hamiltonian in action--angle variables:
\beq{nfcm}
K^{(CM)}(I_y,I_z,\theta_y,\theta_z)=K_0(I_y,I_z)+K_r(I_y,I_z,\theta_y-\theta_z)+R^{(r)}(I_y,I_z,\theta_y,\theta_z)\ ,
\eeq
where $K_0$ depends only on the actions; $K_r$ is the resonant part depending on the actions as well as on
the angles, but just through the combination $\theta_y-\theta_z$, which corresponds to the synchronous resonance;
$R^{(r)}$ represents the reduced remainder function. This procedure leads to have, by construction, that $\dot I_y+\dot I_z=0$
up to the remainder.

\subsection{The indirect method}\label{sec:indirect}
An alternative approach to the normalization (Section~\ref{norm:form}) and center manifold reduction (Section~\ref{sec:center})
has been adopted in \cite{CPS} on the basis of \cite{JM}. It consists in splitting the above procedure in two steps: a preliminary transformation to get a Hamiltonian in which only terms which do not contain integer powers of the product $Q_1 P_1$ are eliminated, and a subsequent resonant normalization. We refer to this procedure as the \sl indirect method. \rm

It can be verified that this procedure coincides with that described above only at the first order in
the resonant term. From the second order on, the two procedures provide different results on which we will comment later on.

\section{Analytical estimates of the bifurcation values}\label{sec:estimates}

In this section we provide a method to give an analytical estimate of the value at which a bifurcation to halo orbits occurs.
We describe in Section~\ref{sec:1ot} the method based on a normalization to first order, while in Section~\ref{sec:second}
we extend the method using a second order normal form. Of course the method can be implemented to higher orders (see Section~\ref{sec:higher});
however, one should keep in mind that Birkhoff's normalization admits an optimal order, which provides the best result. We recall that, for divergent asymptotic series (as the Birkhoff normal form usually is), an optimal order is reached when partial sums of the remainder series have a minimum \cite{gce}. In practice,
it is not trivial to understand which is the best order of normalization, especially when dealing with peculiar cases like that of $L_3$ for small values of the mass ratio of the primaries (see Section~\ref{sec:results}).

\subsection{First-order theory}\label{sec:1ot}
The model problem given in Section \ref{sec:collinear} is characterized by just one parameter, the mass ratio $\mu$. Therefore, all coefficients appearing in the Hamiltonian and its integrable approximation should be expressed in terms of such parameter. The explicit computation of these coefficients is however quite difficult, mostly in view of the following two facts:

1. The distances $\gamma_j$, $j=1,2,3$, of the collinear equilibria from
the closest primary are the solutions of the fifth order Euler's equations and therefore cannot be easily expressed in terms of $\mu$; this difficulty  propagates through the coefficients $c_n(\mu)$ of the expansion in the Hamiltonian \eqref{ham1}.

2. The diagonalizing transformation simplifies the quadratic part of the Hamiltonian, but scrambles its higher-order terms determining complicated expressions for the coefficients in the normal form.

We may try to circumvent the first problem by exploiting a series expansion of the $\gamma_j$ in terms of the mass ratio, by means of which one may hope to get explicit results valid in the limit of a small mass ratio. About the second problem there is little to do, except to compute the normalizing transformation after a generic linear transformation and, only in the end, to substitute the explicit diagonalizing transformation. In  doing so, already the second-order normal form contains coefficients for which an explicit algebraic representation is quite cumbersome. Therefore, in \ref{sec:appendix}, we provide just the expression of the coefficients of the normal form Hamiltonian truncated to the first order. In this way we get explicit analytic first-order formulae which happen to be quite useful, especially for $L_1$ (see Figure~\ref{SL1L2} below and the results presented in Section \ref{Hill:sec}).

To obtain an overall picture of the results, we compute the second-order (and higher-order) normal form only for a finite number of mass ratios in the whole range $0<\mu\le1/2$ by firstly substituting the mass parameter in the original Hamiltonian and, afterwards, by normalizing with numerical coefficients. By proceeding in this semi-analytical way, it is a simpler matter to reach very high orders. This procedure works very well as it is confirmed by the comparison
between some available numerical experiments and the analytic first-order results.
First-order thresholds, being a rough approximation, are obviously less accurate but, by using the quantities of \ref{sec:appendix}, any value of $\mu$ can be chosen and results can be obtained without the need of recomputing the normal form. The only case in which problems appear is that of $L_3$ in the limit of small mass-ratios, see Section~\ref{sec:L3} below.

Truncating the normal form whenever the first resonant terms appear can be considered a \sl first-order \rm resonant perturbation approach. From the order of the resonance generated by \equ{hqr} it is straightforward to check that the odd degree terms in the normal form vanish and that the first non-trivial term is $K^{(NF)}_4$. Therefore, truncating \equ{nfcm} to degree two in the actions leads to the following
first-order normal form:
\beq{Hr}
K^{(CM,1)}(I_y,I_z,\theta_y,\theta_z)=\omega_y I_y +\omega_z I_z + [\alpha I_y^2 + \beta I_z^2+I_yI_z(\sigma+2 \tau \cos(2(\theta_y-\theta_z)))]
\eeq
with suitable coefficients $\alpha$, $\beta$, $\sigma$, $\tau$, whose explicit expressions are given in~\ref{sec:appendix}.

We immediately remark that, if either $I_y$ or $I_z$ vanish, one obtains the nonlinear \sl normal modes\rm, namely
$$
E_y \equiv \omega_y I_y + \alpha I_y^2  \ ,
$$
and
$$
E_z \equiv \omega_z I_z + \beta I_z^2\ .
$$

Denoting by
\beq{sec_int}
\E=I_y+ I_z
\eeq
the conserved quantity (the \sl second integral\rm) resulting from the normalization,
the theory developed in \cite{MP11,PM14,CPS} based on the reduction to a 1-DOF system, allows us to describe the main features of the dynamics provided by \eqref{Hr}.

Here we briefly sketch the essentials of the theory. By reduced 1-DOF system we mean the following. Let us make the change of variables (compare with \equ{sec_int}):
\beqa{coord}
\E&=&I_y+I_z\nonumber\\
\R&=&I_y\nonumber\\
\nu &=&\theta_z\nonumber\\
\psi &=&\theta_y-\theta_z\ .
\eeqa
Then, the Hamiltonian \equ{Hr} is transformed into
\beq{hamER}
K^{(R1)}(\E,\R,\nu,\psi)= \omega_z \E+\delta \R+a \R^2+b \E^2+c \E \R+d(\R^2-\E \R)\cos(2\psi)\ ,
\eeq
where the constants are defined as follows: $a=\alpha+\beta-\sigma$, $b=\beta$, $c=\sigma-2\beta$, $d=-2\tau$.
Hamilton's equations associated to \equ{hamER} take the form
\beqa{ERequ}
\dot{\E}&=&0\nonumber  \\
\dot{\R}&=&2d \R (\R-\E)\sin(2\psi)\nonumber\\
\dot{\nu}&=&\omega_z +2b\E+c\R-d\R\cos(2\psi)\nonumber\\
\dot{\psi}&=&\delta+2a\R+c\E+d(2\R-\E)\cos(2\psi)\ .
\eeqa
Therefore we obtain a 1-DOF system in the phase-plane $(\R,\psi)$, parametrized by $\E$. Its equilibria correspond to periodic orbits of the original 2-DOF system.
%
%
%

From the equation $\dot \R=0$, we obtain a first set of solutions valid for any $\psi\in\torus$: it consists just of the normal modes $\R=0$, $\R=\E$. From the coupled equations  $\dot \R=0, \dot \psi =0$, a second set (the \sl periodic orbits in general position\rm) is given by a solution $\R_i$ associated to the resonant combination $\psi=0,\pi$ and by a solution $\R_{\ell}$ with $\psi=\pm \pi/2$. The equilibrium points $(\R_i,0)$ and $(\R_i,\pi)$
correspond to the so-called  \sl inclined \rm periodic orbits while the equilibrium points $(\R_{\ell},\pm\pi/2)$ correspond to \sl loop \rm orbits. These periodic orbits arise as bifurcations from the normal modes when the following existence conditions are satisfied
$$ 0 \leq \R_i, \R_{\ell} \leq \E.$$
Loops bifurcating from the $\R=\E$ normal mode are precisely what we usually call halo orbits.

%
%

These conditions provide the following constraints for the existence of resonant orbits  bifurcating from the normal modes:
\beq{halobif}
\E\geq \E_{\ell y}\equiv{{\delta}\over {\sigma - 2(\alpha+\tau)}}
\qquad {\rm or}\qquad
\E\geq \E_{\ell z}\equiv {{\delta}\over {2(\beta+\tau)-\sigma}}
\eeq
for the halo family (namely loops, with fixed phase relation $\theta_y-\theta_z=0,\pi$) and
\beq{sec_bif}
\E\geq \E_{i y}\equiv{{\delta} \over {\sigma - 2(\alpha-\tau)}}
\qquad {\rm or}\qquad
\E\geq \E_{i z}\equiv {{\delta}\over {2(\beta-\tau)-\sigma}}
\eeq
for the \sl anti-halo \rm family (that is, the inclined with the fixed phase relation $\theta_y-\theta_z=\pm \pi/2$). The first of \equ{halobif} is just the occurrence of the bifurcation of the halo family from the planar Lyapunov orbit, which becomes unstable. A second bifurcation may occur at the value given by the first of \equ{sec_bif}, when the Lyapunov orbit regains stability. An alternative computation of the thresholds based on Floquet theory
is presented in Section~\ref{sec:Floquet}.

To determine the energy level at which the bifurcation takes place,
we write the threshold value of the integral $\E$ as a power series in $\delta$ and we denote by
$\E_N$ a truncation of the series up to an integer order $N$, say
\beq{eserie}
\E_N=\sum_{k=1}^N C_k\ \delta^k\
\eeq
for suitable real coefficients $C_k$. Notice that, due to the form of \equ{halobif} and \equ{sec_bif}, it is reasonable to start
the series in \equ{eserie} with the first order in $\delta$.
Then, we look for a relation on the bifurcating normal mode between $\E$ and $E$, that is the energy associated to the Hamiltonian \equ{ham1}.
The estimate to first order is simply
$$
E_1= \omega_z \E_1 = \omega_z C_1 {\delta}\ ,
$$
which, coming back to the original coefficients, gives the bifurcation value
\beq{epriord}
E_1=\frac{ \omega_z  \delta}{\sigma-2(\alpha+\tau)}\ .
\eeq

A nice feature of the first-order theory is that, given the coefficients of the normal form \equ{Hr}, all dynamical quantities can be formally explicitly computed. For example, a first order estimate of the frequency of the normal modes (Lyapunov orbits) is given by
\beqano
\kappa_y^{(1)}&=&\omega_y+2 \alpha \E\ ,\nonumber\\
\kappa_z^{(1)}&=&\omega_z+2 \beta \E\ .
\eeqano
Analogously, the computation of the variational frequency of orthogonal perturbations of the normal modes can be performed (see \cite{PM14})
starting from the \sl reduced frequency \rm associated to the 1-DOF dynamics. For the variational frequency of the horizontal normal mode $\kappa_y^{(HNM)}$, we obtain the following expression
\begin{eqnarray}
\kappa_y^{(HNM)}&&=\sqrt{4 \tau^2 \E^2 - [(2\alpha-\sigma)\E+\delta]^2}\nonumber\\
&&=\sqrt{(\E_{i y}-\E)(\E_{\ell y}-\E)} \sqrt{(2\alpha-\sigma)^2-4\tau^2}\ ,
\label{varfreq}
\end{eqnarray}
which gives the {\sl rotation number}
\beq{rnpriord}
\rho=\frac{\kappa_y^{(HNM)}}{ \kappa_y^{(1)}}\ ,\
\eeq
providing information about the stability of the normal perturbation. Comparing \equ{varfreq} with the first expressions in \equ{sec_bif}
and \equ{halobif}, we see that the frequency is real (and that the normal mode, namely the planar Lyapunov orbit, is stable)
for $\E <  \E_{\ell y}$ and for $\E >  \E_{i y}$. In the range $\E_{\ell y} < \E < \E_{i y}$ the inverse of the positive
value of \equ{varfreq} gives the rate of growth of the perturbation.

\subsection{Second order theory}\label{sec:second}
To get a higher order estimate of the bifurcation value, we need an explicit expression of the perturbing function to sixth order.
To this end, we reduce also the term $K^{(NF)}_6$ to the center manifold and we implement perturbation theory.
Again we obtain that the order five does not contribute to the average
as well as to the resonant part, while at order six we obtain the following expression:
\beqa{H6}
K^{(CM,2)}(I_y,I_z,\theta_y,\theta_z)&=&\omega_y I_y +\omega_z I_z +\alpha I_y^2 + \beta I_z^2+I_yI_z(\sigma
+2 \tau \cos(2(\theta_y-\theta_z)))\nonumber\\
&+&\alpha_{3300}I_y^3+\alpha_{0033}I_z^3+\alpha_{1122}I_yI_z^2+\alpha_{2211}I_y^2I_z\nonumber\\
&+&2 I_y I_z [\alpha_{2013} I_z + \alpha_{3102}I_y] \cos (2(\theta_y-\theta_z))
\eeqa
for suitable coefficients $\alpha_{a b c d}$ with $a+b+c+d=6$.

The procedure to get the second order estimate has been illustrated in \cite{CPS}: we remark that the results of Section~\ref{sec:1ot} have the same precision of the equilibria of the linearised system. In the framework of a second order theory, the equilibria result from quadratic equations. Following \cite{Henrard}, rather than computing exact solutions for these equations, it is better to get solutions in terms of series in the detuning, truncated at the same order as the normal form. The second order bifurcation value turns out to be
\beq{secord}
\E_2 = \E_1 -\frac{\alpha_{2211}-3\alpha_{3300}-2 \alpha_{3102}}{(\sigma - 2(\alpha+\tau))^3}\ \delta^2\ ,
\eeq
where $\E_1=\delta/(\sigma-2(\alpha+\tau))$.
To convert this result in terms of the energy, the following second order expression for the energy of the normal mode is necessary:
$$
E = (\omega_z+\delta)\E + \alpha\E^2\ .
$$
Using \equ{secord}, we get the bifurcation energy of the halo at second order as
\beq{esecord}
E_2 = E_1 + \left[\frac{\sigma-\alpha- 2 \tau}{(\sigma-2(\alpha+\tau))^2}-\omega_z \frac{\alpha_{2211}-3\alpha_{3300}-2 \alpha_{3102}}{(\sigma-2(\alpha+\tau))^3}  \right] \delta^2.
\eeq
We remark that the formal structure of \equ{H6} is the same both in the \sl direct \rm normalization method (DM in the following) of Section~\ref{norm:form} and in the \sl indirect \rm method (hereafter IM) adopted in \cite{CPS} (see Section~\ref{sec:indirect}).

\subsection{Higher order theory}\label{sec:higher}
A prediction based on an $N$-th order normal form can be obtained by a suitable extension of the theory (\cite{Henrard,MP14}). The Hamiltonian on the center manifold is of the form
\beq{higher}
\tilde K^{(CM,N)}(P_1Q_1,P_2,P_3,Q_2,Q_3)=\sum_{n=1}^N \tilde K^{(NF)}_{2n}(0,P_2,P_3,Q_2,Q_3)
\eeq
for suitable polynomials $\tilde K^{(NF)}_{2n}$ of degree $2n$ in the variables $(P_2,P_3,Q_2,Q_3)$: the symmetric 1:1 resonant normal form only admits even degree terms.
The equilibria of its reduced counterpart can be found by solving an $N$-th degree algebraic equation. The existence condition turns out in a solution of the form \eqref{eserie}. To convert the result in terms of the energy, an $N$-th order expression for the energy on the normal mode is necessary:
\beq{EE}
E = \sum_{n=1}^N a_n \E^n
\eeq
for suitable real coefficients $a_n$.
Plugging \equ{eserie} into \equ{EE} with $\E=\E_N$, one obtains a series in $\delta$.
Truncating such series to order $N$, we get an approximate value of the bifurcation energy of the halo orbit as
$$
E_N = \sum_{n=1}^N \widehat C_n |\delta|^n \ ,
$$
where the $\widehat C_n$ are rational combinations of the coefficients of the normal form. An optimal order of truncation of the procedure can be obtained by looking for the order giving the best asymptotic convergence of the series.

\subsection{A first-order estimate based on Floquet theory}\label{sec:Floquet}
In this Section we give an alternative derivation of the stability/instability transition of the normal modes
(planar and vertical Lyapunov families) on the basis of the Floquet theory
(see, e.g., \cite{SVM}).
With reference to \equ{Hr}, we start by considering small variations around the planar Lyapunov orbit $I_z=0$. For our purposes,
the easiest way to proceed suggests to employ {\it complex-conjugate} variables in order
to represent small variations around the normal mode. Therefore, instead of the $(I_z,\theta_z)$ variables used before, we introduce
the coordinates $(z,w)$ defined as
\beqano
z &=& \sqrt{2\,I_z} i e^{-i\theta_z}\ , \\
w&=&-\sqrt{2\,I_z} i e^{i\theta_z}\ .
\eeqano
Let
$$
T_y = \frac{2 \pi}{\kappa_y}, \quad \kappa_y\in\real , $$
be the period of the normal mode corresponding to the planar Lyapunov orbit.
The periodic oscillation on the normal mode is given by (see Section~\ref{sec:1ot})
\beq{NMO}
I_y = I\ , \quad \theta_y = \theta = \kappa_y  t\ , \quad \kappa_y = \omega_y+2 \alpha I\ .
\eeq
Since $\omega_y=\omega_z+\delta$, the Hamiltonian \eqref{Hr} can be written as
\beqa{Hrr}
K^{(CM,1)}(z,w,I,\theta)&=&\omega_z \left(I + {{z w}\over 2} \right) + \delta I +\nonumber\\
 &&\left[\alpha I^2 + {\beta\over 4} z^2 w^2 + {\sigma\over 2}
I z w - {\tau\over 2} I \left(z^2 e^{2i\theta} + w^2 e^{-2i\theta}\right)\right]\ .
\eeqa
The equations of motion associated to \equ{Hrr} can be written in almost canonical form as
\beqano
\dot I&=&-{{\partial K^{(CM,1)}}\over {\partial\theta}}\ ,\qquad  \dot \theta={{\partial K^{(CM,1)}}\over {\partial I}}\nonumber\\
\dot z&=&2i {{\partial K^{(CM,1)}}\over {\partial w}}\ ,\qquad \dot w=-2i {{\partial K^{(CM,1)}}\over {\partial z}}\ .
\eeqano
We now exploit the integral of motion
\beq{Eshell}
\E = I + {{z w}\over 2}\eeq
to study the variational dynamics on the \sl energy shell: \rm this is equivalent to the use of the Lagrange multiplier as in
\cite{SVM}. From \eqref{Eshell}, we can substitute $\E- zw/2$ in place of $I$ in \eqref{Hrr} and truncate up to second order in the small quantities $z,w$, thus obtaining the time-periodic 1DOF Hamiltonian
\beqano
K^{(CM,2)}(z,w,\theta;\E)&=&(\omega_z + \delta)\E   + \alpha \E^2 - \left(\delta + (2 \alpha - \sigma) \E \right) {{zw}\over 2}\nonumber\\
&-&{{\tau}\over 2} \E \left(z^2 e^{2i\theta} + w^2 e^{-2i\theta}\right)\ ,
\eeqano
in which $\E$ is considered as a constant parameter. We then pass to \sl corotating coordinates \rm $(Z,W)$ (see \cite{BrSi}) by means of
the transformation
$$
Z=z e^{i\theta}, \quad W=w e^{-i\theta}\ ,
$$
leading to the quadratic Hamiltonian
$$
K(Z,W)=- \left(\delta + (2 \alpha - \sigma) \E \right) {{ZW}\over 2} - {{\tau}\over 2} \E \left(Z^2 + W^2\right)\ ,
$$
where constant terms have been neglected. Introducing the \sl Floquet matrix \rm
\beq{FMO}
F \equiv -i \left( \begin{array}{cc}
                        \delta + (2 \alpha - \sigma) \E & 2 \tau \E  \\
                       - 2 \tau \E & -\delta - (2 \alpha - \sigma) \E \\
                      \end{array}
                    \right)\ ,
\eeq
the canonical equations take the linear form
$$
\left( \begin{array}{c} \dot Z \\ \dot W \\ \end{array} \right) = F \left( \begin{array}{c} Z \\ W \\ \end{array} \right)\ .
$$
The corresponding solution is governed by the \sl fundamental matrix \rm ${\rm exp} (t F)$, so that the solution
for the original complex orbital variations can be written in the form
$$
\left( \begin{array}{c} z(t) \\ w(t) \\ \end{array} \right) = M(t) \left( \begin{array}{c} z(0) \\ w(0) \\ \end{array} \right)\ ,
$$
where, using \eqref{NMO}, it is
\beqano
M(t) &=& \left( \begin{array}{cc}
                        e^{-i\theta} & 0  \\
                       0 & e^{i\theta} \\
                      \end{array}
                    \right) {\rm exp} (t F) \nonumber \\
                    &=& \left( \begin{array}{cc}
                        e^{-i \kappa_y t} \left(\cos \lambda t + \frac{F_{11}}{\lambda} \sin  \lambda t \right) & 0  \\
                       0 & e^{i \kappa_y t} \left(\cos \lambda t - \frac{F_{11}}{\lambda} \sin  \lambda t \right) \\
                      \end{array}
                    \right)\eeqano
with $\pm \lambda$ the eigenvalues of the Floquet matrix $F$.

The \sl monodromy \rm matrix is finally defined as
$$
M(T_y)=   \left( \begin{array}{cc}
                        \cos 2\pi \frac{\lambda}{\kappa_y} + \frac{F_{11}}{\lambda} \sin  2\pi \frac{\lambda}{\kappa_y}  & 0  \\
                       0 & \cos 2\pi \frac{\lambda}{\kappa_y} - \frac{F_{11}}{\lambda} \sin 2\pi \frac{\lambda}{\kappa_y}  \\
                      \end{array}
                    \right)\ .
$$
where we denoted by
$T_y $
the period of the normal mode. The transition from stability to instability or viceversa is given by the condition
$$
{\rm Trace}(M(T_y))=2 \cos 2\pi \frac{\lambda}{\kappa_y} = 2\ ,
$$
which corresponds to $\lambda=0$. An explicit computation
of the eigenvalues $\lambda$ and of the solution $\lambda=0$
using the components of the matrix \eqref{FMO} gives the threshold values
$$
 \E_{\ell y}= {{\delta} \over {\sigma - 2(\alpha+\tau)}}\ , \quad
\E_{i y}     = {{\delta} \over {\sigma - 2(\alpha-\tau)}}\ ,
$$
which coincide with those obtained by examining the conditions for the existence of the critical points of the reduced Hamiltonian (compare with
Section~\ref{sec:1ot}).

By following an analogous procedure, one can analyze the stability transitions of the \sl vertical normal mode: \rm
in this case, the periodic oscillation on the normal mode is given by
$$
I_z = I, \quad \theta_z = \kappa_z t, \quad \kappa_z = \omega_z+2 \beta I
$$
and the Floquet matrix is
$$
F \equiv i \left( \begin{array}{cc}
                        \delta - (2 \beta - \sigma) \E & - 2 \tau \E  \\
                       2 \tau \E & -\delta + (2 \beta - \sigma) \E \\
                      \end{array}
                    \right)\ .
$$
The conditions for vanishing of its eigenvalues now give the threshold values
$$
 \E_{i z} =      {{\delta} \over {2(\beta-\tau)-\sigma}}\ , \quad
 \E_{\ell z} =  {{\delta} \over {2(\beta+\tau)-\sigma}}\ ,
$$
again in agreement with the results of Section~\ref{sec:1ot}.

\section{Results}\label{sec:results}

In this section we present an overview of the results by comparing the analytical estimates obtained in
Section~\ref{sec:estimates} with the numerical values available in the literature (\cite{Henon1,Henon2,GJMS,GM}) or by private communications (\cite{GPC}).
Due to the peculiarity of $L_3$ with respect to the other collinear points, we split the discussion, devoting
Section~\ref{sec:L1L2} to the analysis of the results for $L_1$ and $L_2$, while Section~\ref{sec:L3} provides
the results concerning $L_3$.
A discussion of the small mass limit for $L_1$ and $L_2$ is presented in Section~\ref{Hill:sec}, while small mass ratios for $L_3$ are described in Section~\ref{qkep:sec}.
Precisely, we shall call \sl Hill's case, \rm the case of $L_1$ and $L_2$ when $\mu$ tends to zero, since it is equivalent to let one of the primaries
tend to infinity as in the classical lunar theory studied by G. Hill in \cite{Hill}. When dealing with $L_3$, we shall
refer to the case $\mu\rightarrow 0$ as the \sl quasi--Kepler \rm problem, since it corresponds to a nearly
two--body problem in rotating coordinates (\cite{simo3}).

\subsection{Bifurcation thresholds for $L_1$ and $L_2$}\label{sec:L1L2}
A comparison of the results between the analytical estimates at different orders and the available numerical values
(\cite{Henon1,Henon2,GJMS,GM,GPC}) are displayed in Tables~\ref{tab1:results}--\ref{tab3:results}; in this section we
concentrate on the collinear points $L_1$ (Table~\ref{tab1:results}) and $L_2$ (Table~\ref{tab2:results}),
while Section~\ref{sec:L3} will be devoted to $L_3$.

The numerical evaluation of the bifurcation thresholds in the limit of very small mass-ratio $\mu \to 0$,
the so-called Hill's case,
and that of equal masses ($\mu=1/2$) has been performed by M. H\'enon (\cite{Henon1}) in his seminal works
on the investigation of periodic orbits in the framework of the circular, spatial, restricted three--body problem.
Beside Hill's and the equal masses cases, we consider also two intermediate examples: the barycenter--Sun and the Earth--Moon cases. For these two systems the numerical data have been remarkably obtained in \cite{GJMS,GM,GPC}.
We recall that the barycenter--Sun case is provided by the gravitational attraction between the Earth--Moon barycenter
and the Sun.

As far as the analytical estimates are concerned,
the first-order predictions have been computed by using formula \equ{epriord},
while the second-order predictions are obtained via \eqref{esecord} by using the normal form coefficients
computed with the direct method (DM) and the indirect one (IM); in order to ease the comparison with numerical data, we compute
the values of the \sl physical \rm bifurcation energy, obtained by implementing the conversion formulae (\ref{Ephys1}--\ref{Ephys3}).

We also report the results obtained computing the normal form at higher orders, up to the sixth order (which corresponds to degree 7 in the actions).
The number of digits reported in Tables~\ref{tab1:results} and \ref{tab2:results} is dictated by the asymptotic value
at which an approximate convergence is attained. We stress that the numerical results provided in the last line of Tables 1, 2, 3 are given up to the 5th decimal digit as in \cite{GM}.

\begin{table}[h]
\begin{tabular}{|l|l|l|l|l|}
  \hline
   & Hill's case & barycenter--Sun & Earth--Moon & equal masses \\
  \hline
  & $\mu \to 0$ & $\mu=\mu_{bS}$ & $\mu=\mu_{EM}$ & $\mu=1/2$ \\
  \hline
  \hline
  First order & -1.500000 & -1.500415 & -1.587193 & -1.961675 \\
  Second order (IM) & -1.500000 & -1.500417 & -1.587175 & -1.961535 \\
  Second order (DM) &-1.500000 & -1.500417 & -1.587175 & -1.961534 \\
  Third order (DM) & -1.500000 & -1.500416 & -1.587176 & -1.961536 \\
  Fourth order (DM) & -1.500000 & -1.500416 & -1.587176 & -1.961536 \\
  Fifth order (DM) & -1.500000 & -1.500416 & -1.587176 & -1.961536 \\
  Sixth order (DM) & -1.500000 & -1.500416 & -1.587176 & -1.961536 \\
  \hline
  \hline
  Numerical & -1.50000 & -1.50042 & -1.58718 & -1.96154 \\
  \hline
 \end{tabular}
  \vskip.2in
\caption{Results for the analytical bifurcation estimates for $L_1$ up to a normal form
of order 6 and the numerical values obtained in \cite{Henon1,Henon2,GJMS,GM}, physical energy, see \equ{Ephys1};
the values of the mass ratios are $\mu_{bS}=3.0404326 \times 10^{-6}$ and $\mu_{EM}= 0.01215058$.}
\label{tab1:results}
\end{table}

From the analysis of the data in Table~\ref{tab1:results} we infer that we obtain a very good agreement between
the analytical predictions and the numerical data. The theoretical results improve as the
order of normalization increases, reaching a reasonable agreement with the numerical value (up to the 5th or even the 6th decimal place) at the fifth order
of normalization in the whole range of masses.

\begin{table}[h]
\begin{tabular}{|l|l|l|l|l|}
  \hline
   & Hill's case & barycenter--Sun & Earth--Moon & equal masses \\
  \hline
  & $\mu \to 0$ & $\mu=\mu_{bS}$ & $\mu=\mu_{EM}$ & $\mu=1/2$ \\
  \hline
  \hline
  First order & -1.500000 & -1.500412  &  -1.575838 & -1.524509 \\
  Second order (IM)  & -1.500000 & -1.500413 & -1.576065 & -1.552699 \\
  Second order (DM)  & -1.500000 & -1.500413 & -1.576087 &-1.548191\\
  Third order (DM) & -1.500000& -1.500413 & -1.576055 & -1.543863 \\
  Fourth order (DM) & -1.500000 & -1.500413 & -1.576060  & -1.544834 \\
  Fifth order (DM) & -1.500000 & -1.500413 & -1.576060 & -1.544864 \\
  Sixth order (DM) & -1.500000 & -1.500413& -1.576060 & -1.544820 \\
  \hline
  \hline
  Numerical & -1.50000 & -1.50041 & -1.57606 & -1.54476 \\
  \hline
 \end{tabular}
  \vskip.2in
\caption{Results for the analytical bifurcation estimates for $L_2$ up to a normal form
of order 6 and the numerical values obtained in \cite{Henon1,Henon2,GM,GPC}, physical energy, see \equ{Ephys2};
the values of the mass ratios are $\mu_{bS}=3.0404326 \times 10^{-6}$ and $\mu_{EM}= 0.01215058$.}
\label{tab2:results}
\end{table}

Quite similarly, the values shown in Table~\ref{tab2:results} referring to the collinear point $L_2$
show a good agreement between the analytical and numerical results. Again we find that, with the exception of the
case with $\mu=1/2$ with a discrepancy on the 5th digit, the 5th or 6th decimal digit is typically reached
at the fifth order of normalization.

\begin{figure}[h]
\centering
\includegraphics[scale=0.4]{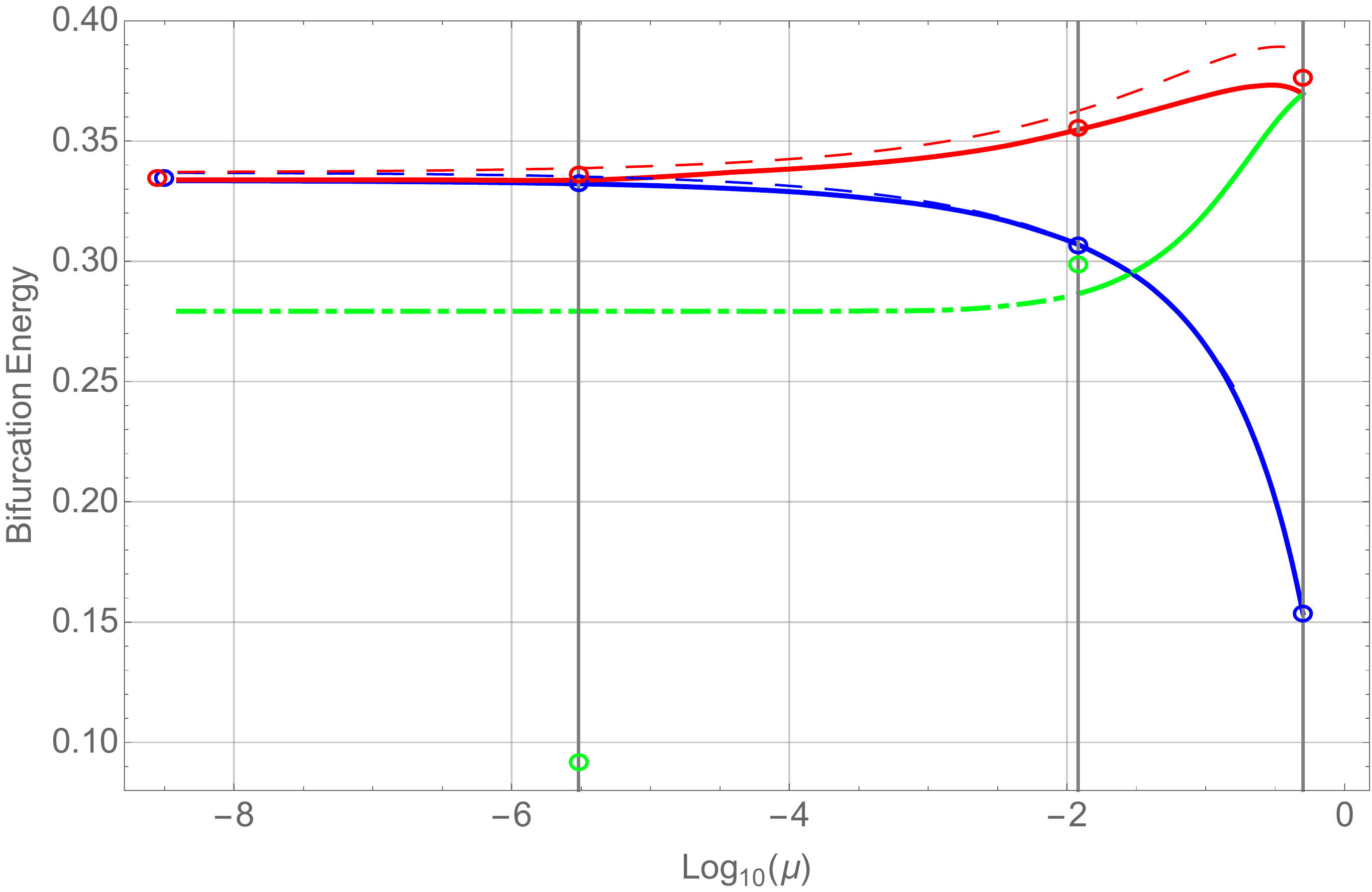}
\caption{Bifurcation thresholds  computed via \eqref{esecord} as a
function of the mass ratio: $L_1$ (blue), $L_2$ (red), $L_3$
(green). The dashed red and blue curves provide the bifurcation thresholds obtained through
\equ{E2serie} and \equ{E1serie} below. The dot-dashed part referring to $L_3$ corresponds
to results obtained when the normal form has already reached the optimal order.}
\label{SL1L2}
\end{figure}

To have a global view, we proceed to compute several values of the bifurcation threshold as a function of the
mass ratio of the primaries and then we interpolate the results. Precisely,
the bifurcation thresholds in the rescaled energy, obtained by \eqref{esecord} using the normal form coefficients
computed with the direct method, are plotted for the whole interval of definition of $\mu$ in Figure~\ref{SL1L2},
where the curves are obtained interpolating 20 points. The blue and red curves refer respectively to $L_1$ and $L_2$: the continuous curves are based on the second-order theory, the dashed ones on the first-order theory (see also Section~\ref{Hill:sec} below).
The vertical lines mark the three reference cases: the barycenter--Sun, the Earth--Moon and the equal-mass values. The circles denote the numerical data reported in the tables.

\begin{figure}
\centering
\includegraphics[scale=0.6]{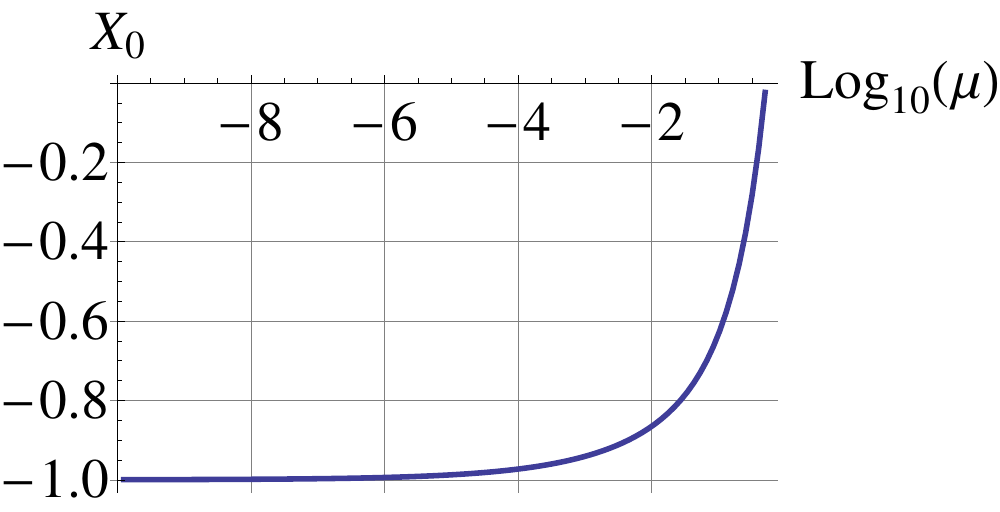}
\includegraphics[scale=0.6]{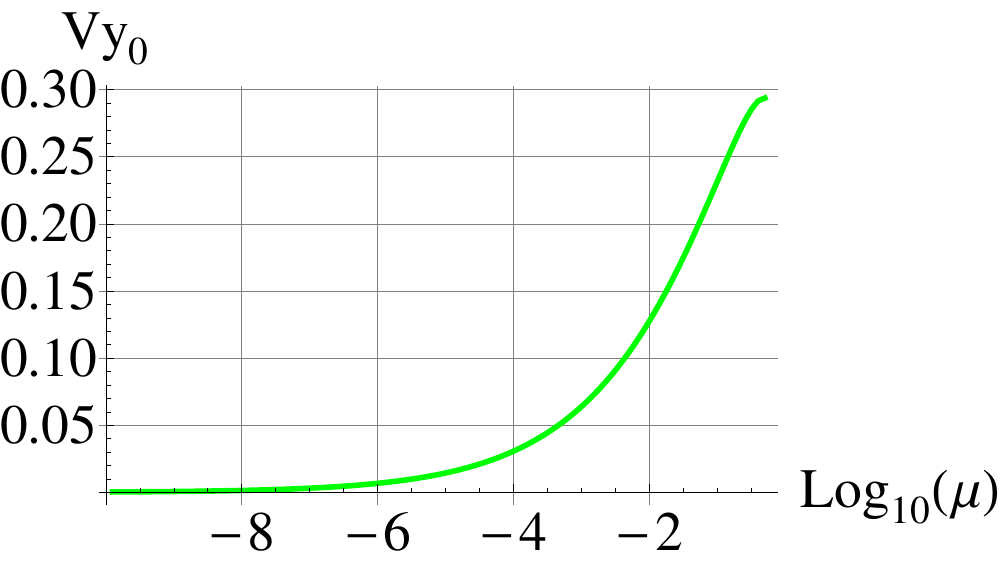}\\
\includegraphics[scale=0.6]{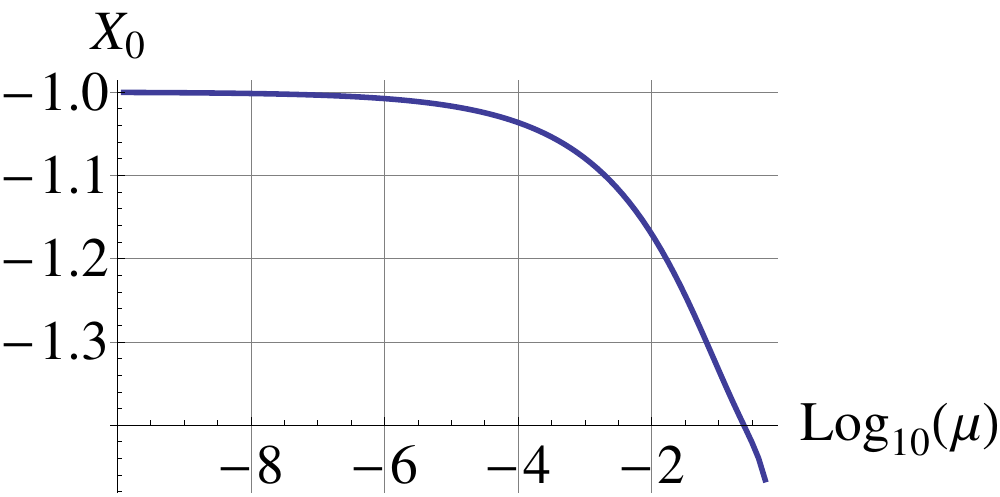}
\includegraphics[scale=0.6]{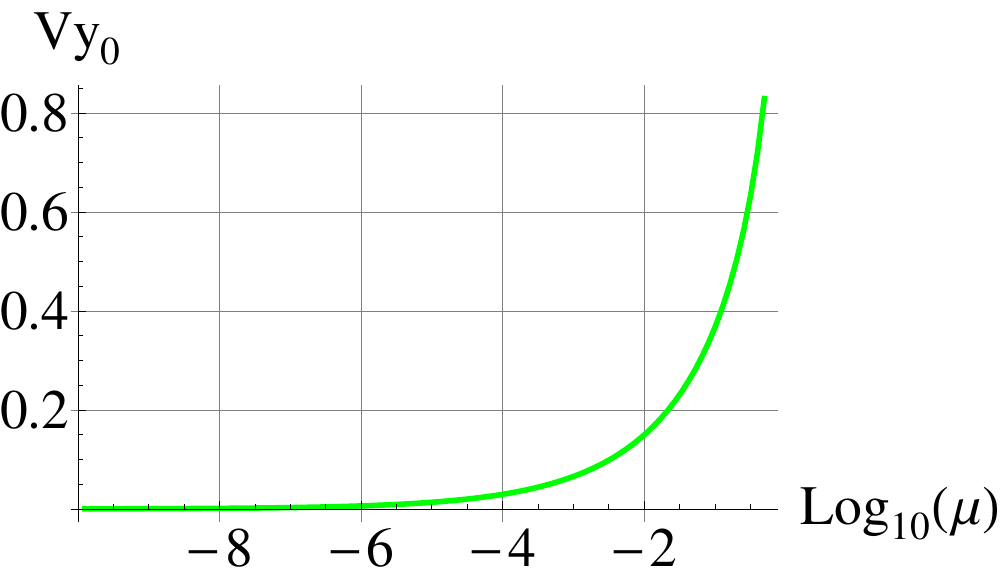}
\caption{Initial values for $X$ (left panels) and $\dot Y$ (right panels)
of the \sl first \rm halo orbits as a function of the mass ratio for $L_1$ (upper panels) and $L_2$ (lower panels).}
\label{X0vy0}
\end{figure}

In view of concrete applications, we find it convenient to give also the initial values of $X_0$, $\dot Y_0$
as a function of $\mu$, which correspond to the bifurcations of the halo orbits. We notice that the plots presented in
Figure~\ref{X0vy0} are in substantial agreement with the values reported in \cite{Henon1,Henon2,Howell}.

\subsection{Bifurcation thresholds for $L_3$}\label{sec:L3}
The dynamics around $L_3$ presents some difficulties, especially in the limit of small masses, in view of the degeneracy
of the problem and the smallness of the perturbation.
Even on the numerical side it is not easy to obtain accurate simulations. In fact, Figure~\ref{SL1L2} provides the
bifurcation threshold of $L_3$ for different values of $\mu$ (green line). However, we need to split the interpolating curve
in two parts: the continuous line (from high values to about $\log_{10}(\mu)=-2$) corresponds to the parameter region in which
the normal form is computed at orders below the optimal one, while
the dot-dashed part of the curve corresponds to results for which the normal form has already reached the optimal order.
We specify that in this framework by \sl optimal order \rm we mean the order at which we get the best agreement
with the numerical threshold. In subsection \ref{sec:convergence} we investigate the problem of the convergence of the normal form obtaining a more rigorous estimate of the optimal order. In the prototypical Earth-Moon case, these results confirm that the second-order evaluation of the threshold is the best at this mass-ratio (with a relative error $\simeq 10^{-2}$) and rapidly deteriorates with decreasing mass. Overall, by looking at Table~\ref{tab3:results} we see how the situation be definitely worse
than for $L_1$ and $L_2$. In fact, reasonable results are obtained only for mass values $\mu > 10^{-2}$, say bigger than the Earth--Moon case, while for smaller masses the approach seems
to fail.
The prediction for the barycenter--Sun case ($-1.40804$, a value kindly provided
in \cite{GPC}) is drastically overestimated.


In our analytical approach the construction of the normal form for $L_3$
is obtained through the same procedure as for the other points $L_1$ and $L_2$.
 It is important to stress that for $L_3$ the indirect method fails
in providing a reliable prediction already at $\mu$ around $0.1$, when it starts to give divergent values of the threshold. The reason for this phenomenon is the following: in the transformation which aims only at eliminating
from the Hamiltonian those terms which do not satisfy the condition to be proportional to $(P_1 Q_1)^k$,
only the terms with \sl small divisors \rm dominated by $\lambda_x$ have an important r\^{o}le.
These terms appear in the generating function already at the first order, thus affecting the normal form
constructed in the indirect way. Since it can be proven (see Section~\ref{Hill:sec}) that  $\lambda_x$ goes to zero
with the square root of $\mu$, there are divergent terms affecting the convergence of the indirect normal form.
These terms are absent by construction in the normal form obtained with the direct method. We will come back on these issues in Section~\ref{qkep:sec} below.

\begin{table}[h]
\begin{tabular}{|l|l|l|l|l|}
  \hline
   & quasi-Kepler & barycenter--Sun & Earth--Moon & equal masses \\
  \hline
  & $\mu \to 0$ & $\mu=\mu_{bS}$ & $\mu=\mu_{EM}$ & $\mu=1/2$ \\
  \hline
  \hline
  First order & -1.178200 & -1.178102 & -1.175384 & -1.524509 \\
  Second order (IM)   & -- & -- & -- & -1.552699 \\
  Second order (DM)   & -1.220215 & -1.219855 & -1.223564 & -1.548191 \\
  Third order (DM) & -- & -- & -1.147760 & -1.543863 \\
  Fourth order (DM) & -- & -- & -1.018562 & -1.544834 \\
  Fifth order (DM) & -- & -- & -1.816723& -1.544864 \\
  Sixth order (DM) & -- & -- & 32.782497 & -1.544820\\
  \hline
  \hline
  Numerical & -- & -1.40804 &-1.21177 & -1.54476 \\
  \hline
 \end{tabular}
  \vskip.2in
\caption{Results for the analytical bifurcation estimates for $L_3$ up to a normal form
of order 6 and the numerical values obtained in \cite{Henon1,GM,GPC}, physical energy, see \equ{Ephys1};
the values of the mass ratios are $\mu_{bS}=3.0404326 \times 10^{-6}$ and $\mu_{EM}= 0.01215058$.}
\label{tab3:results}
\end{table}

\subsection{On the asymptotic convergence of the normal form}\label{sec:convergence}
A relevant question concerns the discussion of the asymptotic convergence of the normal form series.
A first information is given by looking at the Tables~\ref{tab1:results}, \ref{tab2:results}, \ref{tab3:results},
which show a striking agreement of predicted and experimental data in the cases of $L_1$ and $L_2$, while the optimal order
of truncation seems to coincide with the second one in case of $L_3$. As already implemented in \cite{JM},
an estimate of the radius of asymptotic convergence of the normal form series is provided by the classical root and ratio criterions, that we are going to apply as follows.

\begin{figure}
\centering
\includegraphics[scale=0.5]{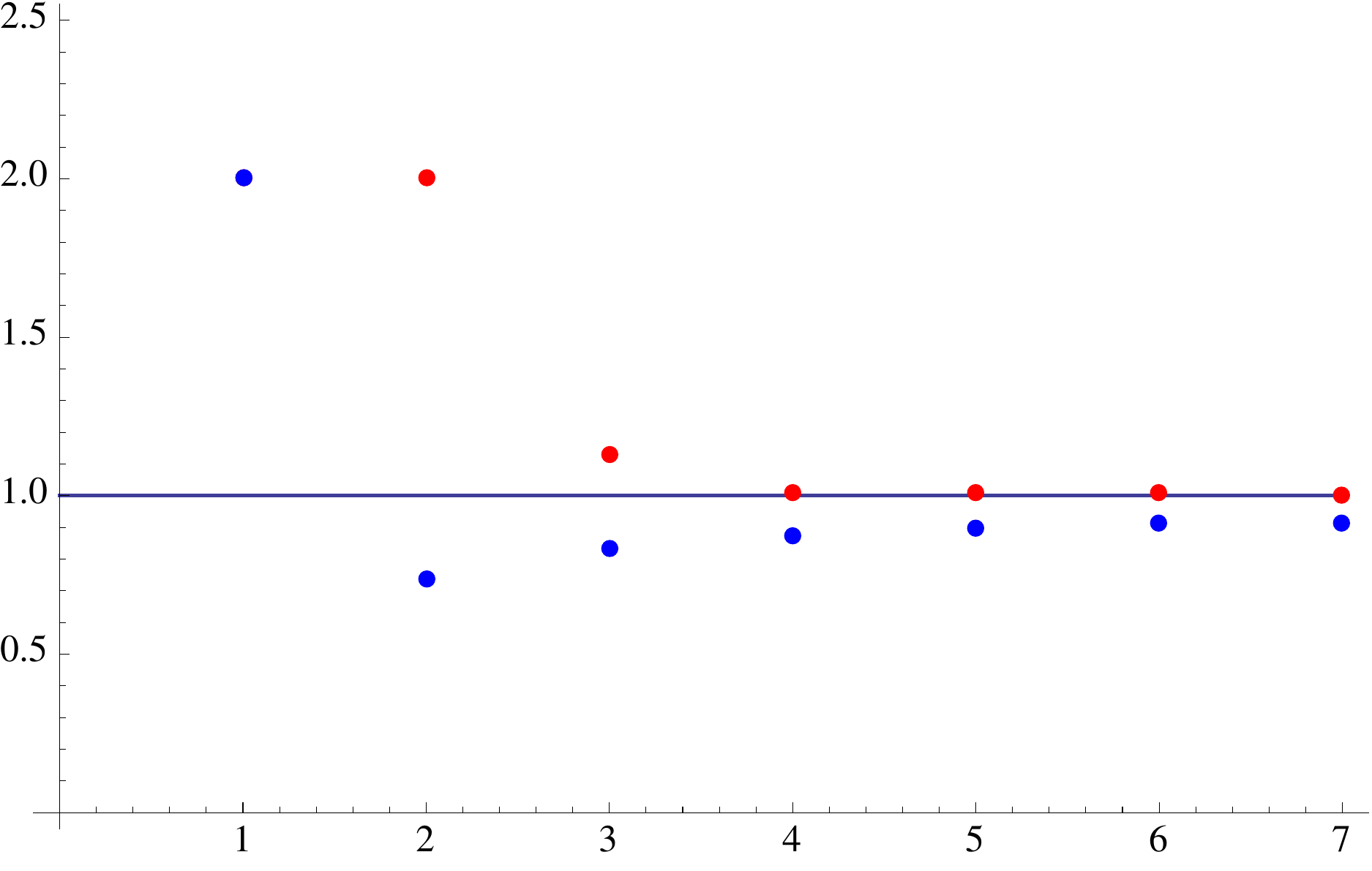}
\includegraphics[scale=0.5]{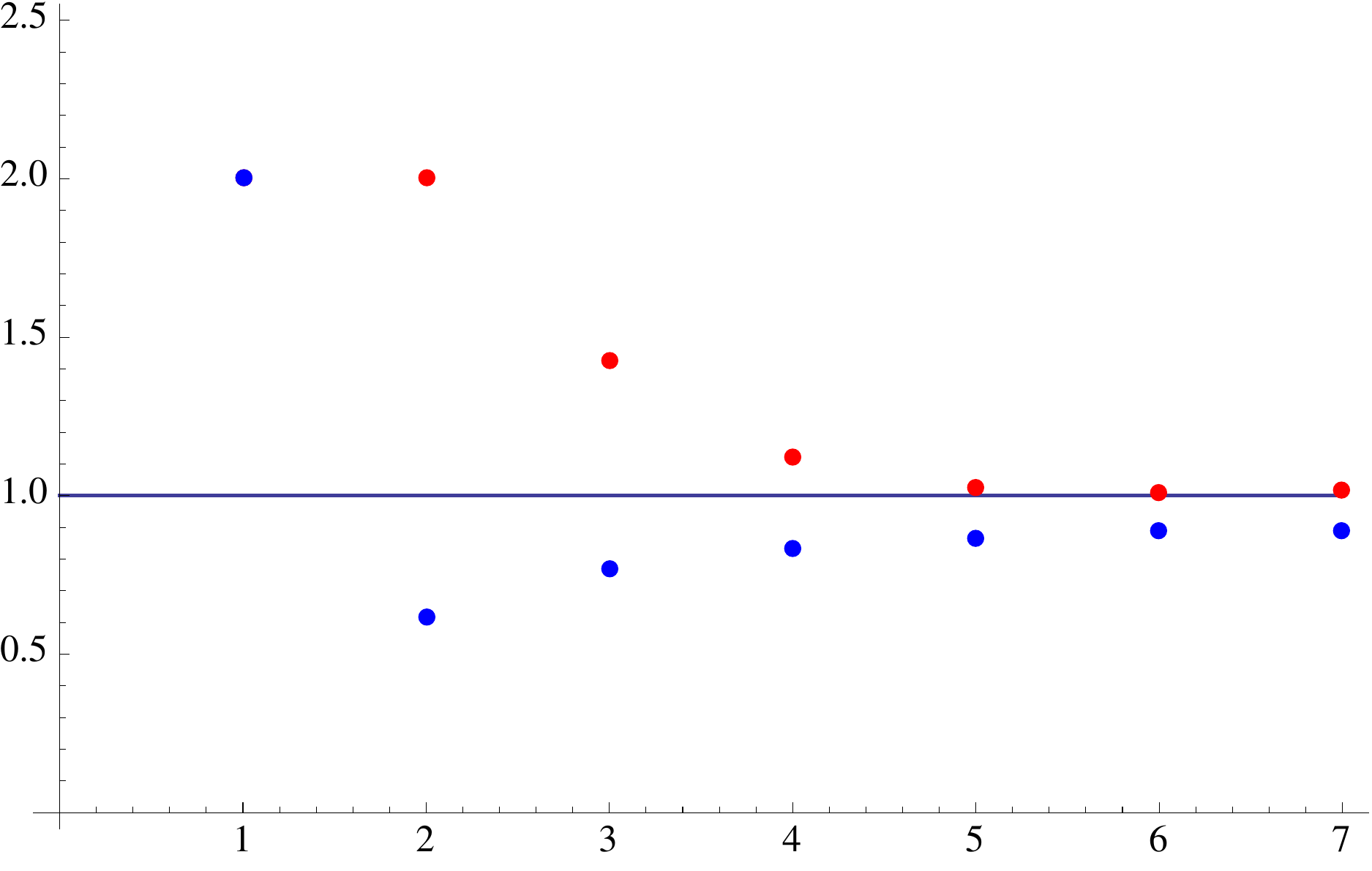}
\includegraphics[scale=0.5]{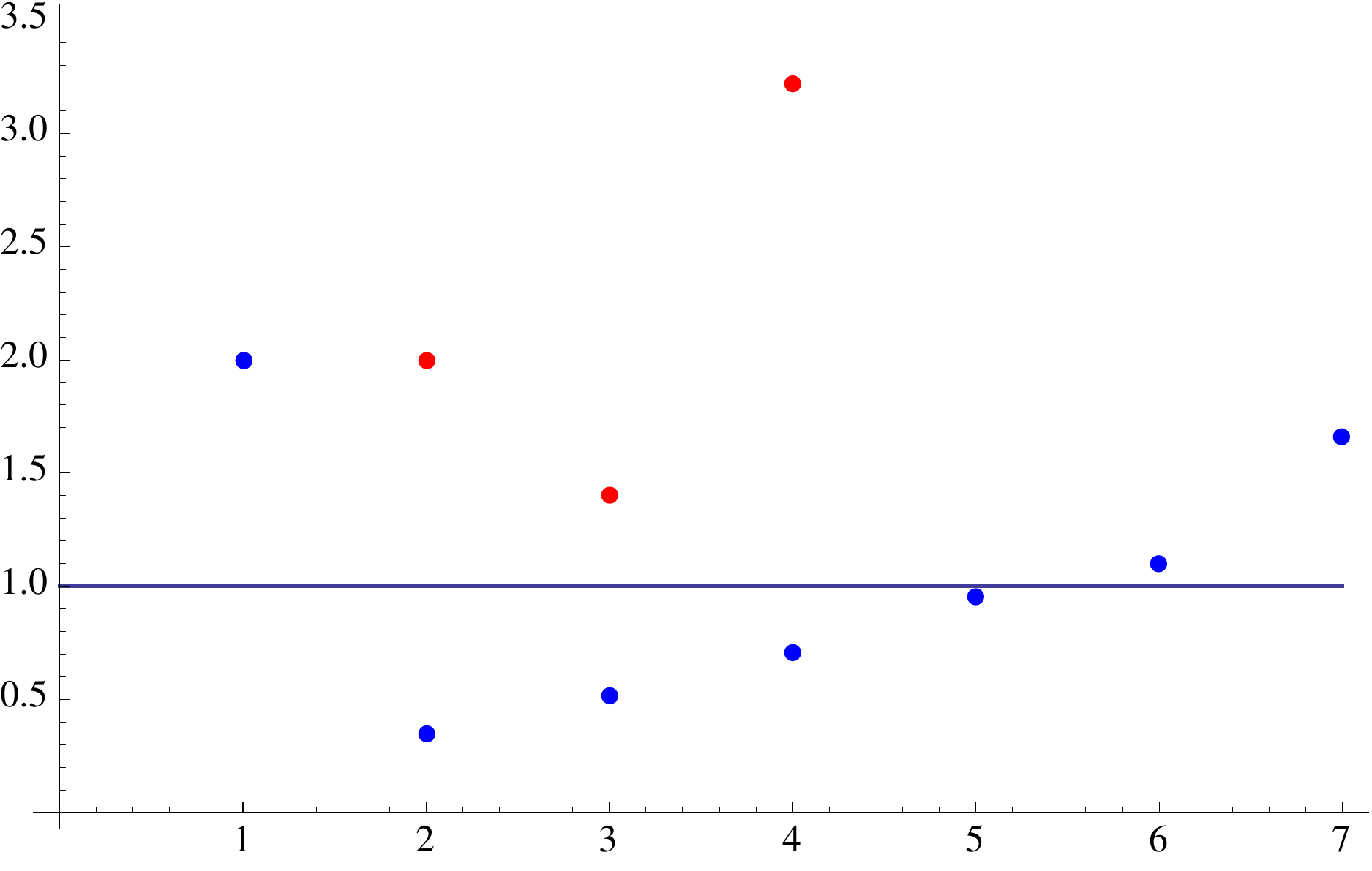}
\caption{Root criterion (blue dots) and ratio criterion (red dots) for the Earth-Moon case for
$L_1$ (upper panel), $L_2$ (middle panel), $L_3$ (lower panel). The abscissa provides the order of normalization,
while the ordinate yieds the quantities in \equ{root} and \equ{ratio}. }
\label{criterion}
\end{figure}

Making reference to the Hamiltonian \equ{higher}, we compute the norm of the
coefficients $\tilde K^{(NF)}_n(0,P_2,P_3,Q_2,Q_3)$. Assuming that such terms $\tilde K^{(NF)}_n$
can be expanded as
$$
\tilde K^{(NF)}_n(0,P_2,P_3,Q_2,Q_3)=\sum_{j_1,j_2,j_3,j_4} h^{(n)}_{j_1,j_2,j_3,j_4} P_2^{j_1}P_3^{j_2}Q_2^{j_3}Q_3^{j_4}
$$
for some complex coefficients $h^{(n)}_{j_1,j_2,j_3,j_4}$, then their norm is defined as
$$
\|\tilde K^{(NF)}_n\|=\sum_{|j_1|+|j_2|+|j_3|+|j_4|=n} |h^{(n)}_{j_1,j_2,j_3,j_4}|\ .
$$
Next, we implement the root criterion by computing the quantity
\beq{root}
\R^{root}_n=\sqrt[n]{\|\tilde K^{(NF)}_n\|}\ .
\eeq
In a similar way, we implement the ratio criterion by computing the quantity
\beq{ratio}
\R^{ratio}_n={{\|\tilde K^{(NF)}_n\|}\over {\|\tilde K^{(NF)}_{n-1}\|}}\ .
\eeq
Figure~\ref{criterion} provides an example of the implementation of the root and ratio criterions
for the Earth--Moon case. We recall that the sixth order normal form corresponds to a computation
up to degree 7 in the actions. For the collinear
points $L_1$ and $L_2$ we have evidence of an agreement with the numerical values which is still improving at order 6. As expected, the case of $L_3$ shows a peculiar behavior in agreement with the
result provided in Table~\ref{tab3:results}, according to which the optimal order of truncation is very low,
possibly coinciding with the second order.

\subsection{Recovering the limit of small masses via the Hill's case: analytic first-order expressions.}\label{Hill:sec}

The case in which the mass ratio tends to zero corresponds, for $L_1,L_2$, to the so-called Hill's case \cite{Hill,Henon2}. We stress that this system, the simplest version of the CSR3BP, has been rigorously proved to be non-integrable \cite{MRSS}.

We recall that we may expand the distances $\gamma_j$, $j=1,2,3$, in series of $\mu$ according to the expressions
\cite{Alebook}:
$$
\gamma_1 = {1\over 3^{1/3}}\left(\frac{\mu}{1-\mu}\right)^{1/3}
- {1\over {3 \times 3^{2/3}}}\left(\frac{\mu}{1-\mu}\right)^{2/3} - \frac1{27} \frac{\mu}{1-\mu} +O\left((\frac{\mu}{1-\mu})^{4/3}\right)
$$
for $L_1$ and
$$
\gamma_2 = {1\over 3^{1/3}}\left(\frac{\mu}{1-\mu}\right)^{1/3}
+ {1\over {3 \times 3^{2/3}}}\left(\frac{\mu}{1-\mu}\right)^{2/3} + \frac1{27} \frac{\mu}{1-\mu} +O\left((\frac{\mu}{1-\mu})^{4/3}\right)
$$
for $L_2$.

For $L_1$ and $L_2$ we get therefore an explicit evaluation of the first-order threshold \eqref{epriord} by using the coefficients reported in \ref{sec:appendix}. For $L_1$ we get the following expressions:

\begin{equation}\label{alfas}
\begin{array}{ll}
\alpha=\alpha_0+\alpha_1\mu^{\frac{1}{3}}+\alpha_2\mu^{\frac{2}{3}}+\alpha_3\mu+O(\mu^{4\over 3})\\
\hspace{1em}=-0.0956176- 0.22769 \mu^{\frac{1}{3}} -0.239225 \mu^{\frac{2}{3}} -0.116094 \mu+O(\mu^{4\over 3})
\end{array}
\end{equation}

\begin{equation}\label{alfas1}
\begin{array}{ll}
\beta=\beta_0+\beta_1\mu^{\frac{1}{3}}+\beta_2\mu^{\frac{2}{3}}+\beta_3\mu+O(\mu^{4\over 3})\\
\hspace{1em}=-0.0775862- 0.230678 \mu^\frac{1}{3} -
 0.245705 \mu^\frac{2}{3}- 0.100155 \mu+O(\mu^{4\over 3})
\end{array}
\end{equation}

\begin{equation}\label{alfas2}
\begin{array}{ll}
\sigma=\sigma_0+\sigma_1\mu^{\frac{1}{3}}+\sigma_2\mu^{\frac{2}{3}}+\sigma_3\mu+O(\mu^{4\over 3})\\
\hspace{1em}=0.0306614 - 0.341373 \mu^\frac{1}{3} - 0.437685 \mu^\frac{2}{3} -
 0.163686 \mu+O(\mu^{4\over 3})
\end{array}
\end{equation}

\begin{equation}\label{alfas3}
\begin{array}{ll}
\tau=\tau_0+\tau_1\mu^{\frac{1}{3}}+\tau_2\mu^{\frac{2}{3}}+\tau_3\mu+O(\mu^{4\over 3})\\
\hspace{1em}=-0.101288- 0.0583243 \mu^\frac{1}{3} -
 0.0245161 \mu^\frac{2}{3} - 0.0254687 \mu+O(\mu^{4\over 3})\ ,
\end{array}
\end{equation}

\begin{equation}\label{alfas4}
\begin{array}{ll}
\delta=\delta_0+\delta_1\mu^{\frac{1}{3}}+\delta_2\mu^{\frac{2}{3}}+\delta_3\mu+O(\mu^{4\over 3})\\
\hspace{1em}=0.0715942- 0.0240373 \mu^{1\over 3} - 0.0150446\mu^{2\over 3}+ 0.0210766 \mu+O(\mu^{4\over 3})\ ,
\end{array}
\end{equation}

\begin{equation}\label{alfas5}
\begin{array}{ll}
\omega_z=\omega_{z_0}+\omega_{z_1}\mu^{\frac{1}{3}}+\omega_{z_2}\mu^{\frac{2}{3}}+\omega_{z_3}\mu+O(\mu^{4\over 3})\\
\hspace{1em}=2+ 1.04004 \mu^{1\over 3} + 0.691078 \mu^{2\over 3} - 0.470486 \mu+O(\mu^{4\over 3})\ ,
\end{array}
\end{equation}
where the coefficients $\alpha_j$, $\beta_j$, $\sigma_j$, $\tau_j$, $\delta_j$, $\omega_{z_j}$, $j=0,1,2,3$, are explicitly listed in \ref{sec:appendix} and
$$
\lambda_x = \sqrt{1 + 2 \sqrt{7}} + \frac{3^{2/3} (7 + 8 \sqrt{7})}{14 \sqrt{1 + 2 \sqrt{7}}} \mu^{1/3}+ ...
$$
By applying \eqref{epriord} we obtain the following expansion in series of the threshold value:
\begin{equation}\label{E1serie}
\begin{array}{ll}
E_1=0.337333 - 0.121141 \mu^\frac{1}{3} - 0.0187564 \mu^\frac{2}{3} -
 0.115146 \mu+O(\mu^{4\over 3})\ .
\end{array}
\end{equation}
For $L_2$ it results:

\begin{equation}\label{alphaL2a}
\begin{array}{ll}
\alpha=\alpha_0+\alpha_1\mu^{\frac{1}{3}}+\alpha_2\mu^{\frac{2}{3}}+\alpha_3\mu+O(\mu^{4\over 3})\\
\hspace{1em}=-0.0956176+ 0.22769 \mu^{\frac{1}{3}} -
 0.239225 \mu^{\frac{2}{3}} + 0.0968215 \mu+O(\mu^{4\over 3})
\end{array}
\end{equation}
\begin{equation}\label{alphaL2b}
\begin{array}{ll}
\beta=\beta_0+\beta_1\mu^{\frac{1}{3}}+\beta_2\mu^{\frac{2}{3}}+\beta_3\mu+O(\mu^{4\over 3})\\
\hspace{1em}=-0.0775862+ 0.230678 \mu^{\frac{1}{3}} -
 0.245705 \mu^{\frac{2}{3}} + 0.114424 \mu+O(\mu^{4\over 3})
\end{array}
\end{equation}

\begin{equation}\label{alphaL2c}
\begin{array}{ll}
\sigma=\sigma_0+\sigma_1\mu^{\frac{1}{3}}+\sigma_2\mu^{\frac{2}{3}}+\sigma_3\mu+O(\mu^{4\over 3})\\
\hspace{1em}=0.0306614 + 0.341373 \mu^{\frac{1}{3}} -0.437685 \mu^{\frac{2}{3}} +
 0.198195 \mu+O(\mu^{4\over 3})
\end{array}
\end{equation}

\begin{equation}\label{gammas}
\begin{array}{ll}
\tau=\tau_0+\tau_1\mu^{\frac{1}{3}}+\tau_2\mu^{\frac{2}{3}}+\tau_3\mu+O(\mu^{4\over 3})\\
\hspace{1em}=-0.101288+ 0.0583243 \mu^{\frac{1}{3}} -
 0.0245161 \mu^{\frac{2}{3}}+ 0.00785379  \mu+O(\mu^{4\over 3})\ ,
\end{array}
\end{equation}

\begin{equation}\label{alphaL2e}
\begin{array}{ll}
\delta=\delta_0+\delta_1\mu^{\frac{1}{3}}+\delta_2\mu^{\frac{2}{3}}+\delta_3\mu+O(\mu^{4\over 3})\\
\hspace{1em}=0.0715942+ 0.0240373 \mu^{1\over 3} -
 0.0150446 \mu^{2\over 3} + 0.0135912 \mu+O(\mu^{4\over 3})\ ,
\end{array}
\end{equation}

\begin{equation}\label{alphaL2f}
\begin{array}{ll}
\omega_z=\omega_{z_0}+\omega_{z_1}\mu^{\frac{1}{3}}+\omega_{z_2}\mu^{\frac{2}{3}}+\omega_{z_3}\mu+O(\mu^{4\over 3})\\
\hspace{1em}=2.- 1.04004 \mu^{1\over 3} + 0.691078 \mu^{2\over 3} -
 1.02951 \mu+O(\mu^{4\over 3})\ ,
\end{array}
\end{equation}
 and
$$
\lambda_x = \sqrt{1 + 2 \sqrt{7}} - \frac{3^{2/3} (7 + 8 \sqrt{7})}{14 \sqrt{1 + 2 \sqrt{7}}} \mu^{1/3}+ ...
$$
which lead to the expansion in series of the threshold value:
\begin{equation}\label{E2serie}
\begin{array}{ll}
E_1=0.337333 + 0.121141  \mu^{\frac{1}{3}}  - 0.0187564  \mu^{\frac{2}{3}}  -
 0.0605633 \mu+O(\mu^{4\over 3})\ .
\end{array}
\end{equation}

Actually the quality of the first-order prediction based on the two series \eqref{E1serie} and \eqref{E2serie} is not limited to small values of the mass ratio. As it can be seen in Figure~\ref{SL1L2} the difference between the bifurcation thresholds  computed via
the second order value provided in \eqref{esecord} (continuous lines) and via the two series \eqref{E1serie} and \eqref{E2serie}
(dashed curves) is quite small in the whole parameter range, especially in the case of $L_1$.

\subsection{The limit of small mass ratio and the Kepler problem: analytic first-order expressions.}\label{qkep:sec}

The expansion of the distance of $L_3$ from the largest primary in the limit of small $\mu$ is given by
$$
\gamma_3 = 1 - \frac7{12} \mu + \frac7{12} \mu^2 - \frac{1127}{20736} \mu^3 + O(\mu^4)\ .
$$
For $L_3$ the coefficients appearing in \equ{Hr} can be expanded as

\begin{equation}\label{alphaL3a}
\begin{array}{ll}
\alpha=\alpha_1\mu+\alpha_2\mu^2+\alpha_3\mu^3+ O(\mu^4)\\
\hspace{1em}=-0.523438 \mu + 5.21802 \mu^2- 43.3717 \mu^3+ O(\mu^4)
\end{array}
\end{equation}

\begin{equation}\label{alphaL3b}
\begin{array}{ll}
\beta=\beta_1\mu+\beta_2\mu^2+\beta_3\mu^3+ O(\mu^4)\\
\hspace{1em}=-0.0175781 \mu + 0.0517578 \mu^2 - 0.0563431 \mu^3+ O(\mu^4)
\end{array}
\end{equation}

\begin{equation}\label{alphaL3c}
\begin{array}{ll}
\sigma=\sigma_2\mu^2+\sigma_3\mu^3+ O(\mu^4)\\
\hspace{1em}=1.53125 \mu^2 - 10.1685 \mu^3+ O(\mu^4)
\end{array}
\end{equation}

\begin{equation}\label{alphaL3d}
\begin{array}{ll}
\tau=\tau_1\mu+\tau_2\mu^2+\tau_3\mu^3+ O(\mu^4)\\
\hspace{1em}=-0.15625 \mu + 0.27417 \mu^2 - 1.27731 \mu^3+ O(\mu^4)\ ,
\end{array}
\end{equation}
\beq{alphaL3e}
\delta=\delta_1 \mu+\delta_2 \mu^2 +\delta_3 \mu^3 +O(\mu^{4})=0.4375 \mu+O(\mu^{2})\ ,
\eeq

\begin{equation}\label{alphaL3f}
\begin{array}{ll}
\omega_z=\omega_{z_0} + \omega_{z_1}\mu+\omega_{z_2} \mu^2 + \omega_{z_3}\mu^3+O(\mu^{4})\\
\hspace{1em}=1 + \, 0.4375 \mu - 1.61784 \mu^2 + 6.75039 \mu^3+O(\mu^{4})\ ,
\end{array}
\end{equation}
where $\alpha_j$, $\beta_j$, $\sigma_j$, $\tau_j$, $\delta_j$, $\omega_{z_j}$ $j=0,1,2,3$, are given in \ref{sec:appendix} and
$$
\lambda_x = \frac12 \sqrt{\frac{21}2} \sqrt{\mu} + O(\mu^{3/2})\ .
$$
The explicit expression of the first-order bifurcation threshold is
$$
E_1=0.321839 + 1.18875\mu - 5.9889 \mu^2 - 108.784 \mu^3 + O(\mu^4)\ .
$$

In the light of these expansions, we have an easy way to interpret the peculiarity of the dynamics around $L_3$ for small masses. In fact, a puzzling issue is the finite value of the threshold (see the green curve of Fig.1) in the limit $\mu \longrightarrow 0$. The reason is that every term in the normal form of degree higher than two \sl vanishes identically
 with the detuning \rm and, since
we have that
$$
{{\delta} \over \omega_z} = \frac7{16} \mu + O(\mu^2)\ ,
$$
they identically vanish in the mass ratio. This degeneracy is due to the fact that, being the dynamics essentially that of the Kepler problem with frequencies $\kappa_y^{(1)} \sim \kappa_z^{(1)} \sim 1+O(\mu)$, perturbed by terms with coefficients
$$
c_n = (-1)^n + C_n \mu
$$
with the $C_n$ numbers of order one, all terms of degree zero in $\mu$ disappear from the normal form \cite{newast}.

We may say that we are now in the framework of a {\sl singular perturbation problem} \cite{BeOr},
that we can define as follows: a perturbation problem in a small parameter ($\mu$ in the present case)
in which the solution of the unperturbed problem (the isotropic harmonic oscillator  in this case,
corresponding to the first-order epicyclic version of the Kepler problem) has qualitative features
distinctly different from those of the exact solution for arbitrarily small, but nonzero values of $\mu$.
In the words of Bender \& Orszag \cite{BeOr}: ``...the exact solution for $\epsilon =0$
($\mu=0$ in our case) is {\sl fundamentally different in character} from the \sl neighboring \rm
solutions obtained in the limit  $\epsilon =0$''. This fundamentally different character
is due to the fact that the unperturbed problem has {\sl only} periodic orbits, whereas the perturbed problem,
for each non-vanishing value of the perturbation parameter, has generically quasi-periodic orbits and isolated
families of periodic orbits triggered by the resonance.
Another factor which enhances the peculiarity of the case of $L_3$ in the limit of small $\mu$ is the difference
with $L_1$ and $L_2$ in the limit $\mu \to 0$: in this case we obtain Hill's problem \cite{Henon2}
and the unperturbed model is now the {\sl anisotropic  harmonic oscillator}, since the detuning term does not vanish
in the limit $\mu \to 0$. We see therefore a clear example of the peculiarity of a case in which a super-integrable system
is perturbed by a non-linear coupling which removes its intrinsic degeneracy, when compared with the more usual case
of a perturbation of a Liouville integrable system which is in general non-degenerate. However, the finite value of the threshold for  $\mu \to 0$ must still be considered a consistent prediction
in the light of the singularity of the perturbation problem. A hint to the reliability of this prediction
comes from the observation that the rotation number \eqref{rnpriord} around the Lyapunov planar orbit tends to zero (see also \cite{simo3}, Section 4.5) and the time-scale of instability
{\sl diverges} in time with a rate exponentially small in $\mu$. These statements are easily verified by using \equ{varfreq}
and observing that every term appearing in the argument of the square root vanishes with $\mu$.

Anyway, the predictions of the bifurcation threshold is given in
Figure~\ref{SL1L2} as the dot-dashed curve, which is based on the
second-order expression \eqref{secord}. With reference to
Table~\ref{tab3:results}, we have good arguments to suspect that
the optimal order of truncation is very small, probably not
greater than the second one (namely, degree six in the
phase-space variables). Therefore, for $L_3$ there should be no
improvement in going to higher orders.

\section{Conclusion}\label{sec:conclusion}
We performed the analysis of the bifurcation sequences of 1:1
resonant Hamiltonian normal forms, which arise from the center
manifold reduction of the collinear points of the circular
restricted 3-body problem. The family of Hamiltonians obtained
with this procedure is composed of perturbations of 2-DOF harmonic oscillators
with slightly different unperturbed frequencies. The harmonic
oscillators can be used as integrable approximations of the
non-integrable dynamical system on the center manifold. The rich
structure of the system on the center manifold can be
investigated with geometric methods, by looking for the existence
and stability of critical points of the reduced 1-DOF system. These
solutions correspond to periodic orbits of the 2-DOF normal form.
In particular, the bifurcation of periodic orbits in general
position from the horizontal normal mode is associated with the
existence of the (stable) family of halo orbits. We compared the
analytical prediction with data obtained from numerical
experiments by computing the value of the energy threshold at
which the bifurcation occurs for arbitrary values of the mass
parameter. An explicit formula for the threshold as a function of
this parameter is obtained starting from the first-order normal
form. A better precision is obtained by normalizing at higher
order for a discrete set of the mass parameter and then interpolating among
the resulting values.

The predictions obtained through our analytical results are
remarkably good for the two collinear points $L_1$ and $L_2$ for
any value of the mass parameter in the range $0<\mu\leq 1/2$. For
what concerns $L_3$, the results are reliable only for values of
the mass parameter greater than about $10^{-2}$
and quickly worsen
with decreasing $\mu$ below this value; in the Earth-Moon case, the second-order
prediction agrees with the numerical datum only at the first
decimal place. The finite value of the threshold for $\mu
\to 0$ must be considered a consistent prediction only in the
light of the singular nature of the perturbation problem. A more effective way to treat this case
most probably needs a radical change of the model to which the perturbation method is applied.\\

%
%
%
%
%
{\bf Acknowledgements.} We thank the anonymous reviewers for comments and suggestions which helped to improve our work. We also thank G. G\'omez and J. M. Mondelo for providing unpublished numerical data and C. Efthymiopoulos, A. Giorgilli and H. Han{\ss}mann for useful discussions.
M.C. was supported by the European MC-ITN grant Astronet-II.
A.C. was partially supported by PRIN-MIUR 2010JJ4KPA$\_$009, GNFM-INdAM and by the
European MC-ITN grant Astronet-II.
G.P. was partially supported by the European MC-ITN grant Stardust and by GNFM-INdAM.


\appendix

\section{}\label{sec:appendix}

For the case of $L_1$ we expand the coefficients of the normal form as in \eqref{alfas}--\eqref{alfas3}
and we obtain the following expressions:

\begin{equation}
\begin{array}{ll}
\alpha_0=\frac{430 - 1561 \sqrt{7}}{38696}\nonumber\\
\\
\alpha_1=\frac{403681129- 710133214 \sqrt{7}}{4492141248 \cdot3^{\frac{1}{3}}}\nonumber\\
\\
\alpha_2=\frac{7615912047925 - 15138513232696 \sqrt{7}}{65185461649728\cdot 3^\frac{2}{3}}\nonumber\\
\\
\alpha_3=\frac{215057379347641787 - 579355824477908807 \sqrt{7}}{11350874807990436096}\nonumber\\
\end{array}
\end{equation}

\begin{equation}
\begin{array}{ll}
\beta_0=-\frac{9}{116}\nonumber\\
\\
\beta_1=-\frac{5969\cdot3^{\frac{2}{3}}}{53824}\nonumber\\
\\
\beta_2=-\frac{1595507}{3121792\cdot3^{\frac{2}{3}}}\nonumber\\
\\
\beta_3=-\frac{54403463}{543191808}\nonumber
\end{array}
\end{equation}

 \begin{equation}
\begin{array}{ll}
\sigma_0=\frac{3}{116}\sqrt{\frac{3}{7}(-383 + 146 \sqrt{7})}\nonumber\\
\\
\sigma_1=-\frac{3^\frac{1}{6} (6769553 - 1082463 \sqrt{7}) \sqrt{163 + 554 \sqrt{7}}}{554508304\nonumber}\\
\\
\sigma_2=-\frac{\sqrt{-(449909592683863331/7) + 46284461373137666 \sqrt{7}}}{458903424\cdot 3^\frac{1}{6}}\nonumber\\
\\
\sigma_3=-\frac{\sqrt{-(13566178178821726882825/7) + (2168079445680944572231 \sqrt{7})/2
 }}{186314790144}\nonumber\\
\end{array}
\end{equation}

\begin{equation}
\begin{array}{ll}
\tau_0=\frac{24876 (-7 + 5 \sqrt{7}) - \sqrt{-690114225129 + 401295726258 \sqrt{7}}}{4488736}\nonumber\\
\\
\tau_1=\frac{-245552521 - 116666072 \sqrt{7} -
 1329626630465352 \sqrt{3} (-9800 + 12771 \sqrt{7})}{292227113679182349696\cdot 3^\frac{5}{6}}\nonumber\\
\\
\tau_2=-\frac{27 (2468130674602081363 - 1313792742465604742 \sqrt{7}) \sqrt{
   1 + 2 \sqrt{7}}}{686797152843693394944\cdot 3^\frac{1}{6}}\nonumber\\
\\
\hspace{1em}-\frac{
  5528 (203 (17991621401387 - 4467226286908 \sqrt{7}) \sqrt{
      85 + 62 \sqrt{7}} +
     77351922 \sqrt{3} (-45133312 + 6427123 \sqrt{7}))}{686797152843693394944\cdot 3^\frac{1}{6}}\nonumber\\
\\
\tau_3=\frac{4279772896 + 13509602135 \sqrt{7}}{652101765504}\nonumber\\
\hspace{2em}-\frac{1}{{6116768064815454196125696}}
\left(-\frac{1831253762265104918553131340897210768586201172585}{21} \right.\\
\hspace{2em}\left.+
 \frac{418836965618388593658800818441465281265686267626  \sqrt{7}}{3}\right)^\frac{1}{2}\nonumber\\
\end{array}
\end{equation}

\begin{equation}
\begin{array}{ll}
\delta_0=-2 + \sqrt{-1 + 2 \sqrt{7}}\nonumber\\
\\
\delta_1=-\frac{\Big(63 (-5 + 7 \sqrt{7}) - \sqrt{42 (8366 + 3367 \sqrt{7})}\Big)^{\frac{1}{3}}}{14^{\frac{2}{3}}}\nonumber\\
\\
\delta_2=-\frac{1127 - 2 \sqrt{-164668 + 177086 \sqrt{7}}}{784\ 3^{\frac{2}{3}}}\nonumber\\
\\
\delta_3=\frac{271}{576} -\frac{ \sqrt{-(\frac{64736369}{21}) + (\frac{12691777 \sqrt{7}}{6})}}{3528}\nonumber\\
\end{array}
\end{equation}

\begin{equation}
\begin{array}{ll}
\omega_{z_0}=2\nonumber\\
\\
\omega_{z_1}=\frac{3^{\frac{2}{3}}}{2}\nonumber\\
\\
\omega_{z_2}=\frac{23}{16\ 3^{\frac{2}{3}}}\nonumber\\
\\
\omega_{z_3}=-\frac{271}{576}\ .\nonumber\\
\end{array}
\end{equation}

For the case of $L_2$, the first few terms of the expansions, see \equ{alphaL2a}--\equ{gammas}, are given by the following expressions:

\begin{equation}
\begin{array}{ll}
\alpha_0=\frac{430 - 1561 \sqrt{7}}{38696}\nonumber\\
\\
\alpha_1=\frac{-403681129 + 710133214 \sqrt{7}}{4492141248 \cdot3^{\frac{1}{3}}}\nonumber\\
\\
\alpha_2=\frac{7615912047925 - 15138513232696 \sqrt{7}}{65185461649728\cdot3^{\frac{2}{3}}}\nonumber\\
\\
\alpha_3=\frac{40926888746031829 + 399917417520057479 \sqrt{7}}{11350874807990436096}\nonumber\\
\end{array}
\end{equation}

\begin{equation}
\begin{array}{ll}
\beta_0=-\frac{9}{116}\nonumber\\
\\
\beta_1=\frac{5969\cdot3^{\frac{2}{3}}}{53824}\nonumber\\
\\
\beta_2=-\frac{1595507}{3121792\cdot3^{\frac{2}{3}}}\nonumber\\
\\
\beta_3=\frac{62154119}{543191808}\nonumber\\
\end{array}
\end{equation}

\begin{equation}
\begin{array}{ll}
\sigma_0=\frac{3}{116}\sqrt{\frac{3}{7}(-383 + 146 \sqrt{7})}\nonumber\\
\\
\sigma_1=-\frac{3^\frac{1}{6} \sqrt{-67904593148 + 38942326934 \sqrt{7}}}{659344}\nonumber\\
\\
\sigma_2=-\frac{(1708763 + 1211194 \sqrt{7}) \sqrt{
 829508697716939878135 + 578984163829071428654 \sqrt{7}}}{138280013612108928 \cdot3^\frac{1}{6}}\nonumber\\
\\
\sigma_3=\frac{\sqrt{14 (-28252063020622261270706 + 17893665708209519504017 \sqrt{7})}}{2608407062016}\nonumber\\
\end{array}
\end{equation}

\begin{equation}
\begin{array}{ll}
\tau_0=\frac{24876 (-7 + 5 \sqrt{7}) - \sqrt{-690114225129 + 401295726258 \sqrt{7}}}{4488736}\nonumber\\
\\
\tau_1=\frac{11459544 \sqrt{3} (-9800 + 12771 \sqrt{7}) - \sqrt{
 7 (514623902740673071495 + 449268874694104189382 \sqrt{7})}}{2518593859712\cdot3^{\frac{5}{6}}}\nonumber\\
\\
\tau_2=-\frac{1998918662848 + 3099921346832 \sqrt{7} }{8478977195601153024\cdot3^{\frac{2}{3}}}\nonumber\\
\\
    \hspace{2em} -\frac{
 3 \sqrt{21 (551490804086039521092038535457867 +
     1332711986256602497930974056496554 \sqrt{7})}}{8478977195601153024\cdot3^{\frac{2}{3}}}\nonumber\\
\\
\tau_3=\frac{372558368 + 290138113 \sqrt{7}}{652101765504}\nonumber\\
\\
\hspace{2em}+\frac{1}{{6116768064815454196125696}}
\left(-\frac{5908980415228982864846540204241770944533372591017}{21} \right.\nonumber\\
\\
\hspace{2em}\left.+
 \frac{320636313928575557216801412344520038042793561962 \sqrt{7}}{3}\right)^\frac{1}{2}\nonumber\\
\end{array}
\end{equation}

\begin{equation}
\begin{array}{ll}
\delta_0=-2 + \sqrt{-1 + 2 \sqrt{7}}\nonumber\\
\\
\delta_1=\frac{\Big(63 (-5 + 7 \sqrt{7}) - \sqrt{42 (8366 + 3367 \sqrt{7})}\Big)^{\frac{1}{3}}}{14^{\frac{2}{3}}}\nonumber\\
\\
\delta_2=\frac{-1127 + 2 \sqrt{-164668 + 177086 \sqrt{7}}}{784\cdot 3^{\frac{2}{3}}}\nonumber\\
\\
\delta_3=\frac{593}{576} -\frac{ \sqrt{-\frac{187748909}{21} + \frac{49407673 \sqrt{7}}{6}}}{3528}\nonumber\\
\end{array}
\end{equation}

\begin{equation}
\begin{array}{ll}
\omega_{z_0}=2\nonumber\\
\\
\omega_{z_1}=-\frac{3^{\frac{2}{3}}}{2}\nonumber\\
\\
\omega_{z_2}=\frac{23}{16\cdot 3^{\frac{2}{3}}}\nonumber\\
\\
\omega_{z_3}=-\frac{593}{576}\ .\nonumber\\
\end{array}
\end{equation}

Finally for $L_3$ the coefficients introduced in \equ{alphaL3a}--\equ{alphaL3d} take the form:
\begin{equation}
\begin{array}{ll}
\alpha_1=-\frac{67}{128}\nonumber\\
\\
\alpha_2=\frac{21373}{4096}\nonumber\\
\\
\alpha_3=\frac{974455051 - 909621504\cdot3^\frac{2}{3}\cdot14^\frac{1}{3} + 299761632\cdot3^\frac{1}{3}\cdot14^\frac{2}{3}}{24772608}\nonumber\\
\end{array}
\end{equation}

\begin{equation}
\begin{array}{ll}
\beta_1=-\frac{9}{512}\nonumber\\
\\
\beta_2=\frac{53}{1024}\nonumber\\
\\
\beta_3=-\frac{7385}{131072}\nonumber\\
\end{array}
\end{equation}

 \begin{equation}
\begin{array}{ll}
\sigma_1=0\nonumber\\
\\
\sigma_2=\frac{49}{32}\nonumber\\
\\
\sigma_3=-\frac{20825}{2048}\nonumber\\
\end{array}
\end{equation}

\begin{equation}
\begin{array}{ll}
\tau_1=-\frac{5}{32}\nonumber\\
\\
\tau_2=\frac{1123}{4096}\nonumber\\
\\
\tau_3=\frac{3876274583 - 1819243008\cdot  3^{\frac{2}{3}}\cdot 14^{\frac{1}{3}} +
  599523264\cdot3^\frac{1}{3}\cdot14^\frac{2}{3}}{173408256}\nonumber
\end{array}
\end{equation}

\begin{equation}
\begin{array}{ll}
\delta_0=0\nonumber\\
\\
\delta_1=\frac{7}{16}\nonumber\\
\\
\delta_2=-\frac{2485}{1536}\nonumber\\
\\
\delta_3=\frac{-951281609 + 454810752\cdot 3^{{\frac{2}{3}}}\cdot 14^{{\frac{1}{3}}} -
 149880816\cdot 3^{{\frac{1}{3}}}\cdot 14^{{\frac{2}{3}}}}{10838016} \\

\end{array}
\end{equation}

\begin{equation}
\begin{array}{ll}
\omega_{z_0}=1\\
\\
\omega_{z_1}=\frac{7}{16}\nonumber\\
\\
\omega_{z_2}=\frac{161}{1536}\nonumber\\
\\
\omega_{z_3}=\frac{-1024017503 + 454810752\cdot 3^{\frac{2}{3}}\cdot 14^{\frac{1}{3}} -
 149880816\cdot 3^{\frac{1}{3}}\cdot 14^{\frac{2}{3}}}{10838016}\ .\nonumber\\
\end{array}
\end{equation}


\vskip.5cm\noindent{\bf References}

\begin{thebibliography}{9}

\bibitem{BeOr}
C. M. Bender, S. A. Orszag, {\sl Advanced Mathematical Methods for Scientists and Engineers I: Asymptotic Methods and Perturbation Theory}, Springer-Verlag,
Berlin (1999)

\bibitem{BrSi}
H. Broer, C. Sim\'o, {\it Resonance Tongues in Hill's Equations: a Geometric Approach}, J. Diff. Eq. {\bf 166}, 290--327 (2000)

\bibitem{BCCP}
S. Bucciarelli, M. Ceccaroni, A. Celletti, G. Pucacco, {\sl Qualitative and analytical results of the bifurcation thresholds to halo orbits},
Annali di Matematica Pura ed Applicata, doi:10.1007/s10231-015-0474-2 (2015)

\bibitem{Alebook}
A. Celletti, {\sl Stability and Chaos in Celestial Mechanics}, Springer-Verlag,
Berlin; published in association with Praxis Publishing Ltd., Chichester, ISBN:
978-3-540-85145-5 (2010)

\bibitem{CPS}
A. Celletti, G. Pucacco, D. Stella, {\sl Lissajous and Halo orbits in the restricted three-body problem},
J. Nonlinear Science {\bf 25}, Issue 2, 343--370 (2015)

\bibitem{conley}
C. C. Conley, \sl Low energy transit orbits in the restricted three-body problem, \rm
SIAM J. Appl. Math. {\bf 16}, n. 4, 732--746 (1968)

\bibitem{gce}
C. Efthymiopoulos, A. Giorgilli, G. Contopoulos, {\it Nonconvergence of formal integrals: II. Improved Estimates for the Optimal Order of Truncation}, Journal of Physics A: Math. Gen. \textbf{37}, 10831--10858 (2004)

\bibitem{ferraz}
S. Ferraz--Mello, \sl Canonical Perturbation Theories, \rm Springer--Verlag (2007)

\bibitem{GLMS}
G. G\'omez, \`A. Llibre, J. Martinez, C. Sim\'o, \sl Dynamics and Mission Design Near Libration Points, Vol. I: Fundamentals: the Case of Collinear Libration Points, \rm World Scientific, Singapore, ISBN: 981-02-4285-9  (2001)

\bibitem{GJMS}
G. G\'omez, \`A. Jorba, J. Masdemont, C. Sim\'o, \sl Dynamics and mission design near libration points. Vol. III:
Advanced methods for collinear points, \rm
World Scientific, Singapore, ISBN: 981-02-4211-5 (2001)

\bibitem{GM}
G. G\'omez, J. M. Mondelo, \sl The dynamics around the collinear equilibrium points of the RTBP, \rm
Physica D {\bf 157}, 283--321 (2001)

\bibitem{GPC}
G. G\'omez, J. M. Mondelo, \sl Private communication \rm (2014)

\bibitem{JM}
\`A. Jorba, J. Masdemont, \sl Dynamics in the center manifold of the collinear points of the restricted three body problem, \rm
Physica D {\bf 132}, 189--213 (1999)

\bibitem{Henon1}
M. H\'enon, \sl Vertical stability of periodic orbits in the restricted problem. I. Equal masses, \rm Astron. \& Astrophys. {\bf 28}, 415--426 (1973)

\bibitem{Henon2}
M. H\'enon, \sl Vertical stability of periodic orbits in the restricted problem. II. Hill's case, \rm Astron. \& Astrophys. {\bf 30}, 317--321 (1974)

\bibitem{Henrard}
J. Henrard, \sl Periodic orbits emanating from a resonant equilibrium, \rm
Celestial Mechanics {\bf 1}, 437--466 (1970)

\bibitem{Hill}
G. W. Hill, \sl Researches in the Lunar Theory, \rm American J. of Mathematics {\bf 1}, n. 1, 5--26 (1878)

\bibitem{Howell}
K. C. Howell, \sl Three-dimensional, periodic, 'halo' orbits, \rm
Celestial Mechanics {\bf 32}, 53--71 (1984)

\bibitem{MP11}
A. Marchesiello, G. Pucacco, \sl Relevance of the 1:1 resonance in galactic dynamics, \rm
Eur. Phys. J. Plus {\bf 126}: 104 (2011)

\bibitem{MP14}
A. Marchesiello, G. Pucacco, {\it Equivariant singularity analysis of the 2:2 resonance}, Nonlinearity, {\bf 27}, 43--66 (2014).

\bibitem{MRSS}
J. J. Morales-Ruiz, C. Sim\'o, S. Simon, \sl Algebraic proof of the non-integrability of Hill's problem, \rm Ergodic Theory and Dynamical Systems, {\bf 25}, 1237--1256 (2005).

\bibitem{MD}
C. D. Murray, S. F. Dermott, {\sl Solar system dynamics},
Cambridge University Press (1999)

\bibitem{newast} G. Pucacco, {\it Normal forms for the epicyclic approximations of the Kepler problem}, New Astronomy {\bf 17}, 475--482 (2012)

\bibitem{PM14}
G. Pucacco, A. Marchesiello, {\it An energy-momentum map for the time-reversal symmetric 1:1 resonance with $\Z_2\times\Z_2$ symmetry}, Physica D \textbf{271}, 10--18 (2014)

\bibitem{richardson}
D. L. Richardson, \sl Analytic construction of periodic orbits about the collinear points, \rm
Celestial Mechanics {\bf 22}, 241--253 (1980)

\bibitem{SVM}
J. A. Sanders, F. Verhulst, J. Murdock,  {\it Averaging Methods in Nonlinear Dynamical Systems}, Springer-Verlag, Berlin (2007)

\bibitem{simo2}
C. Sim\'o, \sl Effective Computations in Celestial Mechanics and Astrodynamics, \rm in ``Modern
Methods of Analytical Mechanics and their Applications", V. V. Rumyantsev and A. V. Karapetyan eds.,
CISM Courses and Lectures 387, 55--102, Springer, Vienna (1998)

\bibitem{simo3}
C. Sim\'o, \sl Dynamical properties in Hamiltonian Systems. Applications to Celestial Mechanics, \rm Lectures delivered at the Centre de Recerca Matem\`atica on January 27--31, 2014: http://www.maia.ub.es/dsg/2014/ $\#1$.

\bibitem{Vf}
F. Verhulst, \sl Discrete symmetric dynamical systems at the main resonances with applications to axi-symmetric galaxies, \rm Royal Society (London), Philosophical Transactions, Series A \textbf{290}, 435--465 (1979)


\end{thebibliography}
\end{document}